\documentclass[prd,preprint,a4paper,superscriptaddress,nofootinbib,11pt]{revtex4}
\usepackage{amsmath,amsfonts,amssymb}
\usepackage{txfonts}
\usepackage[T1]{fontenc}
\usepackage{euscript}
\usepackage{array}
\usepackage{longtable}
\renewcommand{\d}{\mathrm{d}}

\begin{document}


\begin{center}
{\bf  \Large Toward the classification of differential calculi on $\kappa$-Minkowski space and related field theories}
 
 \bigskip
\bigskip

Tajron Juri\'c, \footnote{e-mail: tjuric@irb.hr} Stjepan Meljanac, \footnote{e-mail: meljanac@irb.hr}  Danijel Pikuti\'c, \footnote{e-mail: dpikutic@irb.hr} \\  
Ru\dj er Bo\v skovi\'c Institute, Theoretical Physics Division, Bijeni\v cka  c.54, HR-10002 Zagreb,
Croatia \\[3mm]

Rina \v Strajn, \footnote{e-mail: ri.strajn1@studenti.unica.it}
\\
Dipartimento di Matematica e Informatica,
Universita di Cagliari,\\
 viale Merello 92, I-09123 Cagliari, Italy and INFN, Sezione di Cagliari
 \\[3mm]

\end{center}
\setcounter{page}{1}


\vspace{1.0cm}
{
Classification of differential forms on $\kappa$-Minkowski space, particularly, the classification of all bicovariant differential calculi of classical dimension is presented. By imposing super-Jacobi identities we derive all possible differential algebras compatible with the $\kappa$-Minkowski algebra for time-like, space-like and light-like deformations. Embedding into the super-Heisenberg algebra is constructed using non-commutative (NC) coordinates and one-forms. Particularly, a class of differential calculi with an undeformed exterior derivative and one-forms is considered. Corresponding NC differential calculi are elaborated. Related class of new Drinfeld twists is proposed. It contains twist leading to $\kappa$-Poincar\'e Hopf algebra for light-like deformation. Corresponding super-algebra and deformed super-Hopf algebras, as well as the symmetries of differential algebras are presented and elaborated. Using the NC differential calculus, we analyze NC field theory, modified dispersion relations, and discuss further physical applications.
}



\newpage
\tableofcontents
\newpage

\section{Introduction}
One of the greatest puzzles of modern theoretical physics is that two highly successful physical theories, namely quantum mechanics and general relativity persistently refuse to reconcile with each other, meaning that we still have to wait for a satisfactory theory of quantum gravity. A complementary approach towards quantum gravity is to construct effective theories as an intermediate step between general relativity and a full theory of quantum gravity, and study its physical applications. These results can then be used to unravel some aspects of quantum gravity, particularly those that emerge in the low energy limit and relate them to observable physical phenomena. 

Among many approaches in this direction we chose the approach of noncommutative (NC) gravity and field theory based on deformations of symmetry \cite{1, 2, 3, 4, 5, 6, 7}.  The main idea behind this formalism is to replace the classical symmetries of general relativity (i.e. diffeomorphisms) with a twist deformed Hopf algebra of diffeomorphisms. In a similar way the symmetry of the field theory, which is the Poincar\'e symmetry, is replaced by its twist deformed version. Both symmetry deformations can be interpreted as quantum symmetries and the mathematical framework for their study is the theory of quantum groups and Hopf algebras \cite{8}. 
 
One of the most extensively studied Hopf algebras in the context of physics is the $\kappa$-Poincar\'e algebra \cite{1, 2, 9, 10}  where $\kappa$ denotes the mass-like deformation parameter usually associated with the Planck mass. Following the above idea, application of Einstein's theory of gravity together with the uncertainty principle of quantum mechanics leads to a class of models with spacetime noncommutativity implying that the smooth spacetime geometry of classical general relativity has to be replaced with a noncommutative geometry at distances of the order of the Planck scale \cite{11}.

There are many examples of such geometries including the Groenewald-Moyal plane, $\kappa$-Minkowski space and Snyder space. As a consequence of different approaches to quantum gravity, various phenomenological models have emerged. The most interesting are the Lorentz invariance violation (LIV) and Doubly special relativity (DSR) models \cite{12, 13}. Unlike the LIV models where there exists a preferred reference frame that is singled out, thus manifestly violating the Lorentz invariance, in DSR models the postulates of relativity may be reformulated in such a way as to keep the relativity principle (i.e. equivalence of all inertial observers) intact, while simultaneously allowing the speed of light to depend on its wavelength. One way to do this is through the introduction of yet another invariant parameter (besides that of the speed of light) and by modifying the dispersion relation. 

That the possible existence of signals indicating quantum gravity effects is not merely an academic issue neither a matter of speculation has become evident in astrophysical observations of certain ultra-high energy cosmic rays which seemingly contradict the usual understanding of the high-energy physical processes. These processes include electron-positron production in collisions of high energy photons \cite{14} as well as the observed data from gamma ray bursts \cite{15} which indicates occurrence of time delay between two photons having different energies. It turns out that deviations of this kind can be explained within the DSR framework \cite{16} by modifying the special relativistic dispersion relation with additional terms linear or proportional to some power of the Planck length.

Among quantum symmetries the $\kappa$-Poincar\'e  symmetry is one of the most extensively studied together with the $\kappa$-deformed Minkowski space emerging out of it through the cross product algebra construction \cite{2}. The reason for this is two-fold. First, the quantum field theory with $\kappa$-Poincar\'e  symmetry springs out in a certain limit of quantum gravity coupled to matter fields after integrating out the gravitational/topological degrees of freedom \cite{topology}. This amounts to having an effective theory in the form of a noncommutative field theory on the $\kappa$-deformed Minkowski space. The second reason is that the DSR models are mostly studied within the framework of the $\kappa$-Poincar\'e  algebra, where the $\kappa$-deformed Minkowski spacetime provides the arena for studying the particle kinematics.

An important question arising from the noncommutative nature of  spacetime at the Planck scale is how does this noncommutativity affect the very basic notions of physics such as particle statistics, particularly the spin-statistics relation and how these changes can be implemented into a quantum field theory formalism. One of the approaches to the above question uses the framework of quantum groups \cite{8}, i.e. Hopf algebras and particularly the notion of its quasi-triangular bialgebra structure (i.e. universal $R$-matrix). This structure is important since it is closely related to the corresponding modified algebra of creation and annihilation operators which defines statistics. One could expect that the $\kappa$-deformed oscillator algebra will lead to violation of the Pauli principle, thus indicating the need for reformulating the spin-statistics theorem in the presence of noncommutative geometry \cite{18}.

In formulating field theories on NC spaces, differential calculus plays an essential role. The requirement that this differential calculus is bicovariant \cite{woro} and also covariant under the expected group of symmetries leads to some problems. Regarding the problem of differential calculus on $\kappa$-Minkowski space, Sitarz has shown \cite{19} that in order to obtain bicovariant differential calculus, which is also Lorentz covariant, one has to introduce an extra cotangent direction (calculus has one dimension more than the classical case) for the time-like deformations. While Sitarz considered $3+1$ dimensional space (and developed five dimensional differential calculus), Gonera et al. generalized this work to $n$ dimensions in \cite{20}. In \cite{35c} it was discussed that by gauging this extra one-form, one can introduce gravity in the model, and in \cite{44c} this framework is adapted to formulate field theories. Besides that, the physical interpretation of this extra one-form with no classical analogue remains unclear and one is further motivated to construct a NC differential calculus of classical dimension.

Another attempt to deal with this issue was made in \cite{21} by the Abelian twist deformation of $\mathcal{U}[\mathfrak{igl}(4, \mathbb{R})]$. Bu et al. in \cite{22} extended the  Poincar\'e  algebra with the dilatation operator and constructed a four dimensional differential algebra on the $\kappa$-Minkowski space using a Jordanian twist of the Weyl algebra. Differential algebras of classical dimensions were also constructed in \cite{23} and \cite{24}, from the action of a deformed exterior derivative.

There is a construction \cite{universal} of  a unified graded differential algebra generated by NC coordinates $\hat{x}_{\mu}$, Lorentz generators $M_{\mu\nu}$ and anticommuting NC one-forms $\hat{\xi}_{\mu}$. Enlarging this algebra by introducing a new generator related to the exterior derivative, the authors have found a new unique algebra that satisfies all super-Jacobi identities. This new graded differential algebra is universal, i.e. it is valid for all type of deformations $a_{\mu}$ (time-like, space-like and light-like). When $a_{\mu} = (a_0, ~0)$, the obtained algebra corresponds to the differential algebra in \cite{19} if the extra form $\phi$ is related to the exterior derivative operator $\hat{\d}$. Different realizations of this differential algebra in terms of the super-Heisenberg algebra $\mathcal{SH}$ have been presented, and it has been shown that the volume form \textbf{vol} is undeformed. For light-like deformations $a^2 = 0$  the  $4D$ bicovariant calculus (compatible with \cite{51c}) was derived, along with the corresponding Hopf algebra structure (which is the $\kappa$-Poincar\'e-Hopf algebra), and  the corresponding twist operator. In  \cite{universal} it has been also shown that this twist in $a^2 = 0$ case can be written in a new covariant way and can be expressed in terms of  Poincar\'e  generators only.

We will study the classification of all possible bicovariant differential calculi of classical dimension (the number of one-forms $\hat{\xi}\in\hat{\Omega}^1$ is equal to the number of coordinates $\hat{x}_{\mu}$)  on deformed Minkowski spacetime $\hat{\mathcal{A}}$. More precisely, we propose to investigate algebraic structures underlying differential geometry on the $\kappa$-Minkowski space. As mentioned earlier, differential calculus on this type of space has been studied by several authors \cite{19, 23, 24, 44c, 41}. One way of constructing differential calculus is through the differential graded algebra approach \cite{47} or bicovariant differential calculus on Hopf algebras \cite{48}. Bicovariance condition states that one-forms $\hat{\xi}_{\mu}$ are simultaneously left and right covariant \cite{19, woro}. We will see that the sufficient condition for bicovarince is given by
\begin{equation}\label{bi}
[\hat{\xi}_{\mu},\hat{x}_{\nu}]=iK_{\mu\nu}{}^\alpha \hat{\xi}_{\alpha}, \quad K_{\mu\nu}{}^\alpha\in\mathbb{R},
\end{equation}
 that is, that this commutator is closed in one-forms (and the differential calculus is of classical dimension). Extension of the first-order calculus to the entire differential algebra is not unique which results in different nonequivalent constructions depending on the type of covariance conditions imposed. In this paper we will give all possible bicovariant differential calculi \eqref{bi} that are compatible with $\kappa$-Minkowski algebra and investigate their symmetries. It turns out that there are four types of bicovariant differential calculi that we denote by $\mathcal{C}_1$, $\mathcal{C}_2$, $\mathcal{C}_3$ and $\mathcal{C}_4$. The first three are a one parameter family of algebras and are in general $\mathfrak{igl}(n)$  covariant, while the fourth, that is $\mathcal{C}_4$, is valid only for light-like deformations ($a^2=0$) and is Lorentz and $\kappa$-Poincar\'e covariant. Related new class of new Drinfeld twists is proposed. The $\mathcal{C}_4$ algebra is the only bicovariant differential calculus that is Lorentz covariant. For this case we present a Drinfeld twist and the whole $\kappa$-Poincar\'e-Hopf algebra.

The paper is organized as follows.
In the second section, the bicovariant calculus of classical dimension that is compatible with $\kappa$-Minkowski is analyzed. We present all possible covariant solutions by imposing super-Jacobi identities and the consistency condition for time-like, space-like and light-like deformations.  In section III, we give a realization of differential algebras obtained in the previous section by embedding into the super-Heisenberg algebra $\mathcal{SH}$. In section IV, an algebraic formulation of differential geometry is presented and the formalism for the NC version of the differential calculus is adopted. NC forms are introduced and all (anti-)commutation rules between forms and coordinates are obtained.  In section V, we propose a class of new Drinfeld twists. In section VI, the super-Hopf algebra structure is presented, which enables the investigation of the symmetries of obtained differential algebras. In section VII, field theories in the NC setting are analyzed. Using the setting established in section IV,  the undeformed field theory is analyzed, and after obtaining the Hodge-$*$ operation and integration map, the NC version of free field theories is discussed. We conclude that there is a change in the dispersion relations and final remarks are given in section VIII.


\section{$\kappa$-Minkowski spacetime and the classification of differential calculi}
\subsection{$\kappa$-Minkowski space}
$\kappa$-Minkowski spacetime \cite{1, 2} is usually defined by 
\begin{equation}\label{kappa}
[\hat{x}_{i},\hat{x}_{j}]=0, \quad [\hat{x}_{0},\hat{x}_{i}]=ia_{0}\hat{x}_{i},
\end{equation}
where $a_{0}\propto\frac{1}{\kappa}$ is the deformation parameter usually related to some quantum gravity scale or Planck length \cite{bgmp10, hajume}. Eq. \eqref{kappa} represents the time-like deformations of the usual Minkowski space. We can also look at more general Lie algebraic deformations of Minkowski space   
\begin{equation}\label{kappalie}
[\hat{x}_{\mu},\hat{x}_{\nu}]=iC_{\mu\nu}{}^\lambda \hat{x}_{\lambda},
\end{equation}
where $\hat{x}_{\mu}=(\hat{x}_{0},\hat{x}_{i})$ and structure constants $C_{\mu\nu}{}^\lambda$ satisfy
\begin{equation}
C_{\mu\alpha}{}^\beta C_{\nu\lambda}{}^\alpha+C_{\nu\alpha}{}^\beta C_{\lambda\mu}{}^\alpha +C_{\lambda\alpha}{}^\beta C_{\mu\lambda}{}^\alpha=0.
\end{equation}
\begin{equation}
{C_{\mu\nu}}^\lambda =- {C_{\nu\mu}}^\lambda.
\end{equation}
 For   $C_{\mu\nu}{}^\lambda=a_{\mu}\delta^{\lambda}_{\nu}-a_{\nu}\delta^{\lambda}_{\mu}$ we get
\begin{equation} \label{kappaminkowski}
[\hat x_\mu, \hat x_\nu] = i(a_\mu\hat x_\nu - a_\nu\hat x_\mu).
\end{equation}
For $a_{\mu}=(a_{0},\vec{0})$ we get back to eq. \eqref{kappa} as a special case. Generally, $a_\mu \in M_n$ (Minkowski space $M_n$), $a_\mu=\frac{1}{\kappa}u_\mu$ (where $u^2=-1$ for time-like deformations, $u^2=0$ for light-like deformations and $u^2=1$ for space-like deformations). In this paper we will be working in arbitrary dimensions and Minkowski signature $\eta_{\mu\nu}=\text{diag}(-1,1,...,1)$ (i.e. the spatial eigenvalues of the metric are positive).         

\subsection{Differential calculus of classical dimension}
 We want to construct the most general algebra of differential one-forms $\hat{\xi}_{\mu}\equiv\hat{\d}\hat{x}_{\mu}\in\hat{\Omega}^1$ compatible with $\kappa$-Minkowski spacetime that is bicovariant, i.e. closed in differential forms (the differential calculus is of classical dimension). We define
\begin{equation}\label{forme}
\left\{\hat{\xi}_{\mu},\hat{\xi}_{\nu}\right\}=0,\quad [\hat{\xi}_{\mu},\hat{x}_{\nu}]=iK_{\mu\nu}{}^\alpha \hat{\xi}_{\alpha},
\end{equation}
 where $K_{\mu\nu}{}^\alpha\in\mathbb{R}$ is a tensor with respect to the Lorentz algebra and  generally it is expressed in terms of $a_\mu$ and $\eta_{\mu\nu}$ for $n>2$.\footnote{For $n=2$,  $K_{\mu\nu}{}^\alpha$ is expressed in terms of $a_\mu$, $\eta_{\mu\nu}$ and $\epsilon_{\mu\nu}$.} After imposing super-Jacobi identities the only condition we have for $K_{\mu\nu}{}^\alpha$ comes from  $\left[\hat{x}_{\mu},[\hat{x}_{\nu},\hat{\xi}_{\rho}]\right]+\left[\hat{x}_{\nu},[\hat{\xi}_{\rho},\hat{x}_{\mu}]\right]+\left[\hat{\xi}_{\rho},[\hat{x}_{\mu},\hat{x}_{\nu}]\right]=0$ and it gives
\begin{equation}\label{conditionK}
K_{\lambda\mu}{}^\alpha K_{\alpha\nu}{}^\rho -K_{\lambda\nu}{}^\alpha K_{\alpha\mu}{}^\rho=C_{\mu\nu}{}^\beta K_{\lambda\beta}{}^\rho.
\end{equation}
Other super-Jacobi identities are trivially satisfied by using (\ref{kappalie}) and (\ref{forme}).
Eq. \eqref{conditionK} is valid for general Lie algebraic deformations of spacetime \eqref{kappalie}. We also introduce the  exterior derivative $\hat{\d}\equiv[\hat{\eta},\cdot]$ in a natural way
\begin{equation}\label{exterior}
\hat{\d}\hat{x}_{\mu}=[\hat{\eta}, \hat{x}_{\mu}]=\hat{\xi}_{\mu}, \quad \hat{\eta}^2=0, \quad \hat{\eta}\in \hat{\mathcal{SH}}.
\end{equation} 
When we apply $\hat{\d}=[\hat{\eta},\cdot]$ on \eqref{kappalie} we get
\begin{equation}\label{consist}
[\hat{\xi}_{\mu},\hat{x}_{\nu}]-[\hat{\xi}_{\nu},\hat{x}_{\mu}]=iC_{\mu\nu}{}^\lambda \hat{\xi}_{\lambda}\ \longrightarrow\ K_{\mu\nu}{}^\alpha-K_{\nu\mu}{}^\alpha=C_{\mu\nu}{}^\alpha.
\end{equation}
We  call   eq. \eqref{consist}  the \textsl{consistency condition}.

In order to completely classify differential algebras compatible with $\kappa$-Minkowski space we have to solve \eqref{conditionK} and \eqref{consist} for $K_{\lambda\mu}{}^\alpha$. We demand that $K_{\mu\nu}{}^\alpha\in\mathbb{R}$, that in the limit $a_\mu\rightarrow 0$ the problem reduces to commutative case i.e. $\lim_{a_{\mu}\rightarrow 0} K_{\mu\nu}{}^\alpha=0$ and that the tensor $K_{\mu\nu}{}^\alpha$ has the dimension of length. Therefore, it follows that the most general ansatz (for $n>2$)  is given only in terms of $\eta_{\mu\nu}$ and $a_{\mu}$ via
\begin{equation}
K_{\mu\nu\alpha} = A_0 a_{\mu} a_{\nu} a_{\alpha} + A_1 \eta_{\mu\nu} a_{\alpha} + A_2 \eta_{\mu\alpha} a_{\nu} + A_3 \eta_{\nu\alpha}a_{\mu},
\label{kaminia}
\end{equation}
where $A_1, A_{2}, A_{3}\in\mathbb{R}$ are dimensionless parameters and $A_{0}$ is of dimension (lenght)$^{-2}$, hence $A_0 = \frac c{a^2}$, $c\in\mathbb R$ for $a^2\ne0$ 
and $A_0=0$ for $a^2=0$.
After we impose \eqref{consist} we get
\begin{equation}
A_3=1+A_2.
\end{equation}
Equation \eqref{conditionK} gives
\begin{align}
&A_3(a^2A_0+A_3-1)=0, \\
&A_1(a^2A_0+A_1+1)=0, \\
&A_1A_3a^2=0.
\end{align}
We have four different solutions\footnote{For $n=2$, there are other covariant solutions, for example $K_{\mu\nu\alpha}=\frac{a_\mu a_\nu}{a^2}(c_1a_\alpha+c_2\epsilon_{\alpha\beta}a^\beta)-\eta_{\alpha\mu} a_\nu$, where $c_1, c_2 \in \mathbb R$ are parameters and $a^2\ne0$.}
\begin{enumerate}
\item $A_1=0, ~~~~ A_2=-1, ~~~~ A_3=0, ~~~~ a^2A_0=c$
\item $A_1=0, ~~~~ A_2=-c, ~~~~ A_3=1-c, ~~~~ a^2A_0=c$
\item $A_1=-1-c, ~~~~ A_2=-1, ~~~~ A_3=0, ~~~~ a^2A_0=c$
\item $A_1=-1, ~~~~ A_2=0, ~~~~ A_3=1, ~~~~ a^2=A_0=0$
\end{enumerate}
where $c\in\mathbb{R}$ is a free parameter. We will denote these  four algebras by $\mathcal{C}_{1}$, $\mathcal{C}_{2}$, $\mathcal{C}_{3}$ and $\mathcal{C}_{4}$ respectively\footnote{Where $\mathcal{C}$ stands for $covariant$.}. It is important to note that the first three solutions $\mathcal{C}_{1,2,3}$   are valid for all $a^2\in\mathbb{R}$. There are two cases: when $a^2=0$ then $A_{0}=c=0$, and when $a^2\neq 0$ then $A_{0}=c/a^2$. The fourth solution $\mathcal{C}_{4}$ is only valid in the light-like case $a^2=0$. Explicitly for the tensor $K_{\mu\nu\alpha}$ we have
\begin{equation}\begin{split}\label{cetri}
\mathcal C_1 :&\quad K_{\mu\nu\alpha}= \begin{cases}
\dfrac c{a^2}a_\mu a_\nu a_\alpha - \eta_{\mu\alpha}a_\nu, & \text{if }a^2\ne 0
\\ -  \eta_{\mu\alpha}a_\nu, & \text{if }a^2=0. \end{cases} \\
\mathcal C_2 :&\quad K_{\mu\nu\alpha}=\begin{cases}
\dfrac c{a^2} a_\mu a_\nu a_\alpha-c\eta_{\mu\alpha}a_\nu+(1-c)\eta_{\nu\alpha}a_\mu,& \text{if } a^2\ne 0\\
\eta_{\nu\alpha}a_\mu,& \text{if }a^2=0. \end{cases} \\
\mathcal C_3:&\quad K_{\mu\nu\alpha}=\begin{cases}
\dfrac c{a^2} a_\mu a_\nu a_\alpha - (1+c)\eta_{\mu\nu}a_\alpha-\eta_{\mu\alpha}a_\nu, & \text{if }a^2\ne 0\\
-\eta_{\mu\nu}a_\alpha-\eta_{\mu\alpha}a_\nu, & \text{if }a^2= 0, \end{cases} \\
\mathcal{C}_{4}:&\quad K_{\mu\nu\alpha}=-\eta_{\mu\nu}a_{\alpha}+\eta_{\nu\alpha}a_{\mu}, \quad \text{only for }a^2=0.
\end{split}\end{equation}
Inserting \eqref{cetri} into \eqref{forme} we have 
\begin{equation}\label{comm-xix}\begin{split}
\mathcal C_1: &\quad [\hat\xi_\mu,\hat x_\nu] = \begin{cases}
i\dfrac c{a^2}a_\mu a_\nu (a\hat\xi)-ia_\nu\hat\xi_\mu, & \text{if }a^2\ne0 \\
-ia_\nu\hat\xi_\mu, & \text{if }a^2=0 \end{cases} \\
\mathcal C_2: &\quad [\hat\xi_\mu,\hat x_\nu] = \begin{cases}
i\dfrac c{a^2}a_\mu a_\nu (a\hat\xi)-ica_\nu\hat\xi_\mu+i(1-c)a_\mu\hat\xi_\nu, & \text{if }a^2\ne0 \\
ia_\mu\hat\xi_\nu, & \text{if }a^2=0 \end{cases} \\
\mathcal C_3: &\quad [\hat\xi_\mu,\hat x_\nu] = \begin{cases}
i\dfrac c{a^2}a_\mu a_\nu(a\hat\xi)-i(1+c)\eta_{\mu\nu}(a\hat\xi)-ia_\nu\hat\xi_\mu, & \text{if }a^2\ne0 \\
-i\eta_{\mu\nu}(a\hat\xi)-ia_\nu\hat\xi_\mu, & \text{if }a^2=0 \end{cases} \\
\mathcal C_4: &\quad [\hat\xi_\mu,\hat x_\nu] = -i\eta_{\mu\nu}(a\hat{\xi})+ia_{\mu}\hat{\xi}_{\nu}, \quad a^2=0 
\end{split}\end{equation}
where we used $a\hat{\xi}\equiv a_{\alpha}\hat{\xi}^{\alpha}$.


There are also some special cases when $c=0$ and $A_0=0$ in \eqref{cetri}. We denote these three special cases by\footnote{Where $\mathcal{S}$ stands for $special$.} $\mathcal S_1$,  $\mathcal S_2$ and $\mathcal S_3$
\begin{equation}\begin{split}
&\mathcal{S}_{1}:  \quad [\hat{\xi}_{\mu},\hat{x}_{\nu}]=-ia_{\nu}\hat{\xi}_{\mu}, \quad a^2 \in \mathbb R \\
&\mathcal{S}_{2}: \quad [\hat{\xi}_{\mu},\hat{x}_{\nu}]=ia_{\mu}\hat{\xi}_{\nu}, \quad a^2 \in \mathbb R \\
&\mathcal{S}_{3}: \quad [\hat{\xi}_{\mu},\hat{x}_{\nu}]=-i\eta_{\mu\nu}(a\hat{\xi})-ia_{\nu}\hat{\xi}_{\mu}, \quad a^2 \in \mathbb R
\end{split}\end{equation}
In section VI we will relate the algebra $\mathcal{S}_{1}$ to \textsl{right covariant} realization, the algebra  $\mathcal{S}_{2}$ to \textsl{left covariant} realization, the algebra $\mathcal{S}_{3}$ to \textsl{Magueijo-Smolin} realization and $\mathcal{C}_{4}$ to \textsl{natural} realization (see \cite{Meljanac-3, ms06, ms11, 90} for more details on realizations). 
It is important to note that the algebra $\mathcal{C}_{4}$ is valid only for light-like deformations $a^2=0$ and is equivalent to the algebra obtained in \cite{universal}.



Solutions $\mathcal C_1$, $\mathcal C_2$, $\mathcal C_3$ are new and each of them corresponds to time-like, light-like and space-like deformation parameter $a_\mu$. For time-like deformation $a_\mu=(a_0,0,0,...)$, differential algebras $\mathcal D_1$, $\mathcal D_2$ were constructed in \cite{41}, 
while $\mathcal D_3$ obtained from $\mathcal C_3$ is a new solution. For $c=1$ we see that algebras  $\mathcal{D}_{1}^{c=1}$ and $\mathcal{D}_{2}^{c=1}$ coincide. This case was in detail investigated in \cite{EPJC}. In \cite{oeckl}  (see Corollary 5.1.) the cases $\mathcal{D}_{1}^{c=0}$ and $\mathcal{D}_{2}^{c=0}$ were obtained from a different construction. For light-like deformation $a^2=0$, we have also found three new solutions (see eq. \eqref{comm-xix} for $a^2=0$). The important properties related to twisted super Hopf algebras are presented in section VI.

\textbf{Remark}

The solution $\mathcal C_4$ holds only for light-like deformation $a^2=0$. This solution was constructed first in \cite{universal}. In \cite{universal}, we have constructed universal $\kappa$-Poincar\'e covariant differential calculus over $\kappa$-Minkowski space. This universal algebra has been generated by $ \left\{\hat x_\mu, ~ M_{\mu\nu}, ~ \hat\eta, ~ \hat\xi_\mu\right\}$ (where $M_{\mu\nu}$ are Lorentz generators) for time-like, light-like and space-like deformation parameter $a_\mu$. If $a^2\ne0$, bicovariant calculus implies that, besides one-forms $\hat\xi_\mu$, there is an additional one-form proportional to $\hat\eta$. Only in case of light-like deformation $a^2=0$, the one-forms have classical dimension, i.e. the additional one-form does not appear. However, in present paper we start with $\kappa$-Minkowski space and demand classical dimension for one-forms. Among all the solutions for light-like deformation ($a^2=0$), only solution $\mathcal C_4$ coincides with light-like case constructed in \cite{universal}. It follows that solution $\mathcal C_4$ is compatible with $\kappa$-Poincare Hopf algebra, which is demonstrated in sections V A and VI C.


\section{Embedding into  the super-Heisenberg algebra}
\subsection{Realizations via super-Heisenberg algebra}
We can enlarge the $\kappa$-Minkowski and differential algebra by introducing  derivatives  $\partial_{\mu}$ (we could also use the physical momenta $p_{\mu}=-i\partial_{\mu}$) and Grassmann derivative $q_{\mu}$ with the following properties
\begin{equation}\label{part}
[\partial_{\mu},\partial_{\nu}]=\left\{q_{\mu},q_{\nu}\right\}=[\partial_{\mu},q_{\nu}]=0
\end{equation}
Then the most general algebra between NC coordinates, forms and derivatives  that satisfies super-Jacobi identities is given by
\begin{equation}\begin{split}\label{deformed}
&[\partial_{\mu}, \hat{x}_{\nu}]=\varphi_{\mu\nu}(\partial), \quad [\partial_{\mu}, \hat{\xi}_{\nu}]=0,\\
&[q_{\mu},\hat{x}_{\nu}]=q^{\beta}\tilde{\varphi}_{\mu\beta\nu}(\partial), \quad \left\{q_{\mu},\hat{\xi}_{\nu}\right\}=\upsilon_{\mu\nu}(\partial)
\end{split}\end{equation}
where $\varphi_{\mu\nu}$ and $\upsilon_{\mu\nu}$ are invertible matrices satisfying  following differential equations
\begin{equation}\begin{split}\label{eq}
&\frac{\partial \varphi^\sigma{}_\mu}{\partial(\partial^\alpha)}\varphi^\alpha{}_\nu-\frac{\partial \varphi^\sigma{}_\nu}{\partial(\partial^\alpha)}\varphi^\alpha{}_\mu=iC_{\mu\nu}{}^\alpha \varphi^\sigma{}_\alpha\\
&\frac{\partial\tilde\varphi^{\alpha\beta}{}_\mu}{\partial (\partial^\rho)}\varphi^\rho{}_\nu-\frac{\partial\tilde\varphi^{\alpha\beta}{}_\nu}{\partial (\partial^\rho)}\varphi^\rho{}_\mu+\tilde\varphi^\alpha{}_{\rho\mu}\tilde\varphi^{\rho\beta}{}_\nu-\tilde\varphi_\rho{}^\beta{}_\mu\tilde\varphi^{\alpha\rho}{}_\nu=iC_{\mu\nu}{}^\lambda \tilde\varphi^{\alpha\beta}{}_\lambda\\
&\frac{\partial \upsilon^\rho{}_\mu}{\partial(\partial^\sigma)}\varphi^\sigma{}_\nu-\tilde\varphi^\rho{}_{\sigma\nu}\upsilon^\sigma{}_\mu=iK_{\mu\nu}{}^\alpha\upsilon^\rho{}_\alpha
\end{split}\end{equation}
where for $a_{\mu}\rightarrow 0$ we have the undeformed super-Heisenberg algebra, that is $\varphi_{\mu\nu}=\eta_{\mu\nu}$, $\tilde{\varphi}_{\alpha\beta\gamma}=0$, $\upsilon_{\mu\nu}=\eta_{\mu\nu}$ and $K_{\alpha\beta\gamma}=0$.
The above eqs. \eqref{eq} are complicated and have infinitely many solutions, but we can analyze some special cases. If we take that $\tilde{\varphi}_{\mu\nu\rho}=0$ then the third equation yields
$\frac{\partial \upsilon^{\rho}_{\ \mu}}{\partial(\partial^{\sigma})}\varphi^{\sigma}_{\ \nu}=iK^{\ \ \  \alpha}_{\mu\nu}\upsilon^{\rho}_{\ \alpha}$, which for fixed $\varphi$ and $K$ determines the unique deformation of differential forms, i.e. $\upsilon_{\mu\nu}\neq\eta_{\mu\nu}$. On the other hand, if we take that the algebra of differential forms is undeformed $\left\{q_{\mu},\hat{\xi}_{\nu}\right\}=\eta_{\mu\nu}$, that is $\upsilon_{\mu\nu}=\eta_{\mu\nu}$ then we have a unique solution for $\tilde{\varphi}^{\rho}_{\ \mu\nu}=-iK^{\ \ \ \rho}_{\mu\nu}$. In both cases the function $\varphi$ is determined only by the first relation in eq. \eqref{eq} and the general solutions are analyzed in \cite{Meljanac-4}.

The exterior derivative $\hat{\text{d}}$  introduced in \eqref{exterior} can be written as
\begin{equation}\label{d}
\hat{\text d}=[\hat{\eta},\cdot],
\end{equation}
where the generator $\hat{\eta}\in\hat{\mathcal{SH}}$ is given by
\begin{equation}
\hat{\eta}=\hat\xi_\alpha\partial^\alpha.
\end{equation}
The definition of exterior derivative \eqref{exterior} and \eqref{d} fixes the realization $\varphi_{\mu\nu}(\partial)$ and we have
\begin{equation}\label{linear}
\varphi^\rho{}_\mu=\delta^\rho_\mu-iK_{\alpha\mu}{}^\rho\partial^\alpha.
\end{equation}
Since $K_{\mu\nu\alpha}$ is determined by \eqref{cetri}, we see that for fixed algebra of differential forms and NC coordinates we have a unique linear realization for $\hat{x}_{\mu}$ given by \eqref{linear}.

We can embed the whole algebra \eqref{kappa}, \eqref{forme}, \eqref{part}, \eqref{deformed} into the super-Heisenberg algebra defined by 
\begin{equation}\begin{split}\label{SH}
&[x_{\mu},x_{\nu}]=[\partial_{\mu},\partial_{\nu}]=0,\quad [\partial_{\mu},x_{\nu}]=\eta_{\mu\nu}, \\
&\{\xi_{\mu},\xi_{\nu}\}=\{q_{\mu},q_{\nu}\}=0, \quad  \{\xi_{\mu},q_{\nu}\}=\eta_{\mu\nu},\\
&[x_{\mu},\xi_{\nu}]=[x_{\mu},q_{\nu}]=[\partial_{\mu},\xi_{\nu}]=[\partial_{\mu},q_{\nu}]=0,\\
\end{split}\end{equation}
and find the realizations for all NC generators in terms of the generators of super-Heisenberg algebra via
\begin{equation}\label{realizacija}
\hat x_\mu=x_\alpha\varphi^\alpha{}_\mu+\xi_\alpha\tilde\varphi^{\alpha\beta}{}_\mu q_\beta, \quad \hat\xi_\mu=\xi_\alpha\upsilon^\alpha{}_\mu.
\end{equation}
It is straightforward to verify equations \eqref{deformed}-\eqref{realizacija} with given realization \eqref{realizacija}.  In \cite{algebroid} we have analyzed the realization of NC coordinates by embedding the whole algebra into the Heisenberg algebra and there we have outlined the relations between different realizations (that is all the bases of Heisenberg algebra) by similarity transformations. Similarly, we can relate two realizations $\left\{x^\prime, \partial^\prime, \xi^{\prime}, q^{\prime}\right\}$ and $\left\{x, \partial, \xi, q\right\}$ of super-Heisenberg algebra
\begin{equation}\begin{split} 
x^{\prime}_\mu&=\mathcal E x_\mu \mathcal E^{-1}=x_\alpha\psi^\alpha{}_\mu(\partial)+\xi_\alpha\tilde\psi^{\alpha\beta}{}_\mu (\partial) q_\beta\\
\partial^{\prime}_\mu&=\mathcal E \partial_\mu \mathcal E^{-1}=\Lambda_\mu(\partial)\\
\xi^{\prime}_\mu&=\mathcal E \xi_\mu \mathcal E^{-1}=\xi_\alpha \omega^\alpha{}_\mu(\partial)\\
q^{\prime}_\mu&=\mathcal E q_\mu \mathcal E^{-1}=q^\alpha\tilde\omega_{\mu\alpha}(\partial).
\end{split}\end{equation}
Hence, if we have one solution of  \eqref{eq} we can generate another solution via similarity transformations $\mathcal{E}$. 

\subsection{Class of differential calculi with undeformed exterior derivative and 1-forms}

In the special case when $\upsilon_{\mu\nu}=\eta_{\mu\nu}$ we have  $\tilde{\varphi}^{\rho}_{\ \mu\nu}=-iK^{\ \ \rho}_{\mu\nu}$ which leads to
\begin{equation}\label{real}
\hat x_\mu=x_\mu-iK_{\alpha\mu}{}^\rho(x_\rho\partial^\alpha+\xi_\rho q^\alpha)=x_\mu-iK_{\alpha\mu}{}^\rho L_\rho{}^\alpha, \quad \hat\xi_\mu=\xi_\mu, \quad \hat\eta=\xi_\alpha\partial^\alpha,
\end{equation}
where $L_{\mu\nu}$ generates the $\mathfrak{gl}(n)$ algebra (see Section V for further discussion).
Note that, because $x_\mu^\dagger=x_\mu$, $\partial_\mu^\dagger=-\partial_\mu$, $q_\mu^\dagger=q_\mu$, $\xi_\mu^\dagger=\xi_\mu$ and $[\partial_\mu,x_\nu]=\{\xi_\mu,q_\nu\}=\eta_{\mu\nu}$, it follows that $L_{\mu\nu}$ is antihermitian, so the realization \eqref{linear} is hermitian.

We have only four  classes of covariant realizations for NC coordinates $\hat x_\mu$
\begin{equation}\begin{split}\label{crealizacije}
\mathcal{C}_1\!:& ~ \hat{x}_{\mu}=x_{\mu}-i
\begin{cases}\dfrac c{a^2} a_{\mu}[(ax)(a\partial)+(a\xi)(aq)]+a_{\mu}[(x\partial)+(\xi q)],&a^2\ne0\\ 
a_{\mu}[(x\partial)+(\xi q)],&a^2=0 \end{cases} \\
\mathcal{C}_2\!:& ~ \hat{x}_{\mu}=x_{\mu}-i
\begin{cases}
\dfrac c{a^2} a_{\mu}[(ax)(a\partial)+(a\xi)(aq)]+ca_{\mu}[(x\partial)+(\xi q)]-(1-c)[x_{\mu}(a\partial) +\xi_{\mu}(a q)],&a^2\ne0 \\
[-x_{\mu}(a\partial) -\xi_{\mu}(a q)],&a^2=0
\end{cases}\\
\mathcal{C}_3:& ~ \hat{x}_{\mu}=x_{\mu}-i
\begin{cases}
\dfrac c{a^2} a_{\mu}[(ax)(a\partial)+(a\xi)(aq)]+(1+c)[(a x)\partial_{\mu}+(a\xi)q_{\mu}]+a_{\mu}[(x\partial)+(\xi q)],&a^2\ne0\\
[(a x)\partial_{\mu}+(a\xi)q_{\mu}]+a_{\mu}[(x\partial)+(\xi q)],&a^2=0
\end{cases} \\
\mathcal{C}_4\!:& ~ \hat{x}_{\mu}=x_{\mu}+i[(ax)\partial_{\mu}+(a\xi)]q_{\mu}-i[x_{\mu}(a\partial) +\xi_{\mu}(a q)]=x_{\mu}+ia^{\alpha}M_{\alpha\mu}, \quad a^2=0,\\
\end{split}\end{equation}
where $M_{\mu\nu}=L_{\mu\nu}-L_{\nu\mu}$ are Lorentz generators (see Section V).
It is important to note that \eqref{crealizacije} are all possible covariant linear realizations of \eqref{kappalie}. Namely, if we take  \eqref{real} as an ansatz and just insert it in \eqref{kappa} or \eqref{kappalie}, we would get that the tensor $K_{\mu\nu\alpha}$ has to satisfy \eqref{conditionK} and \eqref{consist} (without refering to differential forms $\hat{\xi}$ or exterior derivative $\hat\d$) \cite{linreal}. In other words only for linear realizations of $\kappa$-Minkowski space, one can have undeformed differential forms $\hat{\xi}_{\mu}=\xi_{\mu}$.

\subsection{Action of $\hat{\mathcal{SH}}$ on $\hat{\mathcal{SA}}$}
Let us mention the
$\kappa$-deformed super-Heisenberg algebra $\hat{\mathcal{SH}}$ which is generated by NC coordinates $\hat{x}_{\mu}$, one forms $\hat{\xi}_{\mu}$,  derivatives $\partial_{\mu}$ and Grassmann derivatives $q_{\mu}$, and $\hat{\mathcal{SA}}\subset\hat{\mathcal{SH}}$ which is a unital subalgebra generated by NC coordinates $\hat{x}_{\mu}$ and one-forms $\hat{\xi}_{\mu}$. 
We can define  the action (for more details see \cite{algebroid, EPJC, kovacevic-meljanac}) 
$\blacktriangleright\  : \hat{\cal SH}\otimes\hat{\mathcal{SA}}\mapsto\hat{\cal SA}$, where, symbolically,  $\hat{\cal SH}=\hat{\mathcal{SA}}\mathcal{ST}$,  and $\cal ST$ is a unital subalgebra of $\hat{\cal H}$ generated by $\partial_{\mu}$ and $q_{\mu}$: 
\begin{equation}\begin{split}\label{crnodjelovanje}
\hat{x}_{\mu} \blacktriangleright \hat{g}(\hat{x},\hat{\xi})&=\hat{x}_{\mu}\hat{g}(\hat{x},\hat{\xi}),\quad \partial_{\mu}\blacktriangleright 1=0,  \\
\hat{\xi}_{\mu}\blacktriangleright\hat{g}(\hat{x},\hat{\xi})&=\hat{\xi}_{\mu} \hat{g}(\hat{x},\hat{\xi}),\quad q_{\mu}\blacktriangleright 1=0,\\
\partial_{\mu}\blacktriangleright \hat{x}_{\nu}&=\eta_{\mu\nu}, \quad q_{\mu}\blacktriangleright \hat{\xi}_{\nu}=\eta_{\mu\nu}, \\
\partial_\mu \blacktriangleright \hat\xi_\nu &= 0, \quad q_\mu \blacktriangleright \hat x_\nu = 0
\end{split}\end{equation}



\section{NC differential calculi over $\kappa$-Minkowski space}

In section II we have found all possible differential algebras of classical dimension that are compatible with $\kappa$-Minkowski space. Our goal is to develop a differential calculus, that is, to define  the exterior derivative and higher order forms. First we will outline the usual undeformed differential calculus over undeformed Minkowski space, but using our algebraic approach. Then  the NC analog is developed. 

Let us first analyze the usual undeformed differential calculus in the more algebraic language.
In the usual, undeformed differential geometry we would have a unital Abelian algebra $\mathcal{A}$ generated by $x_{\mu}$ and algebra $\Omega^p\subset\mathcal{SA}$, $p\in\{0,1,...,n\}$, generated by $p$-forms $\omega=\omega_{\alpha_1 ... \alpha_p}\xi^{\alpha_1}...\xi^{\alpha_p}$, where $\omega_{\alpha_1 ... \alpha_p}\in\mathcal{A}$ and $\Omega^0 \equiv \mathcal{A}$. Then the exterior derivative is a map $\d: \Omega^p\rightarrow\Omega^{p+1}$ that satisfies the Leibniz rule 
\begin{equation}
\d(fg)=(\d f)g+f(\d g),
\end{equation}
 where $f,g\in\mathcal{A}$ and is simply realized by $\d\equiv[\eta,\cdot]$ and $\eta\equiv\xi^{\alpha}\partial_{\alpha}$. Of course, since $[x_{\mu},\xi_{\nu}]=0$ we have that one forms are given by 
\begin{equation}
\d f=\xi^{\alpha}(\partial_{\alpha}\triangleright f)=(\partial_{\alpha}\triangleright f)\xi^{\alpha},
\end{equation}
where $\partial_{\alpha}\triangleright f=\frac{\partial f}{\partial x^{\alpha}}$. Usually one forgets about $\triangleright$ and simply writes $\d f=\frac{\partial f}{\partial x^{\alpha}}\d x^{\alpha}$.

Now we analyze the noncommutative case.
Let $\hat{\mathcal{A}}$ be a unital algebra generated by $\hat{x}_{\mu}$ and $\hat{\Omega}^p\subset\hat{\mathcal{SA}}$ an algebra generated by NC p-forms $\hat{\omega}=\hat{\omega}_{\alpha_1 ... \alpha_p}\hat{\xi}^{\alpha_1}...\hat{\xi}^{\alpha_p}$, where $\hat{\omega}_{\alpha_1 ... \alpha_p}\in\hat{\mathcal{A}}$ and $\hat{\Omega}^0 \equiv \hat{\mathcal{A}}$. The exterior derivative $\hat{\d}$ is a map $\hat{\d}: \hat{\Omega}^p\rightarrow\hat{\Omega}^{p+1}$ that satisfies the Leibniz rule\
\begin{equation}\label{leibniz}
\hat{\d}(\hat{f}\hat{g})=(\hat{\d}\hat{f})\hat{g}+\hat{f}(\hat{\d}\hat{g}),
\end{equation}
where $\hat{f}$, $\hat{g}$ $\in\hat{\mathcal{A}}$. Eq. \eqref{leibniz} is easily fulfilled by choosing $\hat{\d}\equiv[\hat{\eta}, \cdot]$ and $\hat{\eta}\equiv\hat{\xi}^{\alpha}\partial_{\alpha}$ as mentioned in the previous sections (see \eqref{exterior}, \eqref{d}). The commutation relations between differential forms $\hat{\xi}_{\mu}$ and an arbitrary element of $\hat{\mathcal{A}}$ can be written as
\begin{equation}\label{lambda}
\hat{\xi}_{\mu}\hat{f}=(\Lambda_{\mu\alpha}\blacktriangleright\hat{f})\hat{\xi}^{\alpha}, \quad \hat{f}\hat{\xi}_{\mu}=\hat{\xi}^{\alpha}(\Lambda^{-1}_{\mu\alpha}\blacktriangleright\hat{f}),
\end{equation}
where $\Lambda_{\mu\nu}$ is expressed in terms of derivatives $\partial_\mu$ and $\Lambda^{-1}_{\mu\nu}\equiv(\Lambda^{-1})_{\mu\nu}$ denotes the inverse matrix, i.e. $\Lambda^{-1}_{\mu\alpha}\Lambda^{\alpha}{}_\nu=\eta_{\mu\nu}$. Since $\hat{\d}\hat{f}=[\hat{\eta},\hat{f}]\in\hat{\Omega}^1\subset\hat{\mathcal{SA}}$ it follows
\begin{equation}\begin{split}
\hat{\d}\hat{f}=[\hat{\eta},\hat{f}]\blacktriangleright 1=\hat{\eta}\hat{f}\blacktriangleright 1=\hat{\eta}\blacktriangleright\hat{f}= \hat{\xi}^{\alpha}(\partial_{\alpha}\blacktriangleright\hat{f}).
\end{split}\end{equation}
That is we have 
\begin{equation}\label{df}
\hat\d\hat f=\hat\xi^\alpha(\partial_\alpha\blacktriangleright\hat f)
=(\partial_\beta\Lambda^\beta{}_\alpha\blacktriangleright\hat f)\hat\xi^\alpha.
\end{equation}
Furthermore, eq. \eqref{df} and Leibniz rule \eqref{leibniz} imply
\begin{equation}\begin{split}\label{racun}
\hat{\d}(\hat{f}\hat{g})&\overset{\eqref{leibniz}}{\equiv}(\hat{\d}\hat{f})\hat{g}+\hat{f}(\hat{\d}\hat{g})\overset{\eqref{df}}{\equiv}\hat{\xi}^{\alpha}\partial_{\alpha}\blacktriangleright(\hat{f}\hat{g})\\
&\overset{\eqref{df}}{=}\hat{\xi}^{\alpha}(\partial_{\alpha}\blacktriangleright \hat{f})\hat{g}+\hat{f}\hat{\xi}^{\alpha}(\partial_{\alpha}\blacktriangleright \hat{g})\\
&\overset{\eqref{lambda}}{=}\hat{\xi}^{\alpha}[(\partial_{\alpha}\blacktriangleright\hat{f})\hat{g}+(\Lambda^{-1}_{\beta\alpha}\blacktriangleright\hat{f})(\partial^{\beta}\blacktriangleright\hat{g})]
\end{split}\end{equation}
and by comparing the first and last line in \eqref{racun} we get the Leibniz rule for $\partial_{\alpha}$ 
\begin{equation}
\partial_{\alpha}\blacktriangleright(\hat{f}\hat{g})=(\partial_{\alpha}\blacktriangleright\hat{f})\hat{g}+(\Lambda^{-1}_{\beta\alpha}\blacktriangleright\hat{f})(\partial^{\beta}\blacktriangleright\hat{g}).
\end{equation}
The Leibniz rule and coproduct are related via $\partial_{\alpha}\blacktriangleright(\hat{f}\hat{g})=m[\Delta\partial_{\alpha}\blacktriangleright(\hat{f}\otimes\hat{g})]$, where $m(a\otimes b)=ab$, so that the coproduct for $\partial_{\mu}$ is given by
\begin{equation}\label{unicop}
\Delta \partial_{\mu}=\partial_{\mu}\otimes 1+\Lambda^{-1}_{\alpha\mu}\otimes\partial^{\alpha},
\end{equation}
and since $\partial_{\mu}$ generates a Hopf algebra of translations it follows that its antipode and counit are
\begin{equation}
S(\partial_{\mu})=-\partial_{\alpha}\Lambda^{\alpha}{}_{\mu}, \quad \epsilon(\partial_{\mu})=0.
\end{equation}
The associativity of the product between forms and elements of $\hat{\mathcal{A}}$ gives
\begin{equation}\begin{split}
\hat{\xi}_{\mu}(\hat{f}\hat{g})&=(\hat{\xi}_{\mu}\hat{f})\hat{g}\\
\overset{\eqref{lambda}}{\Longrightarrow}[\Lambda_{\mu\alpha}\blacktriangleright(\hat{f}\hat{g})]\hat{\xi}^{\alpha}&=(\Lambda_{\mu\alpha}\blacktriangleright\hat{f})\hat{\xi}^{\alpha}\hat{g}\overset{\eqref{lambda}}{\Longleftarrow}\\
&\overset{\eqref{lambda}}{=}(\Lambda_{\mu\alpha}\blacktriangleright\hat{f})(\Lambda^{\alpha}_{\beta}\blacktriangleright\hat{g})\hat{\xi}^{\beta}.
\end{split}\end{equation}
We can read out the Leibniz rule for $\Lambda_{\mu\alpha}$ as 
\begin{equation}
\Lambda_{\mu\alpha}\blacktriangleright(\hat{f}\hat{g})=(\Lambda_{\mu\beta}\blacktriangleright\hat{f})(\Lambda^{\beta}_{\ \alpha}\blacktriangleright\hat{g}),
\end{equation}
and extract the following coproduct
\begin{equation}
\Delta\Lambda_{\mu\nu}=\Lambda_{\mu\alpha}\otimes \Lambda^\alpha{}_\nu.
\end{equation}
Additionally we have
\begin{equation}
S(\Lambda_{\mu\nu})=\Lambda^{-1}_{\mu\nu}, \quad \epsilon(\Lambda_{\mu\nu})=\eta_{\mu\nu}.
\end{equation}

It is important to note that the constructed differential calculus is bicovariant, meaning that all products of higher forms are again differential forms regardless of the way they are multiplied. To illustrate this more closely let us consider two different forms: $\hat{\omega}\in\hat{\Omega}^q$ and $\hat{\theta}\in\hat{\Omega}^p$. Written in basis we have
\begin{equation}\begin{split}
\hat{\omega}&=\hat{\omega}_{\mu_1 ... \mu_q}\hat{\xi}^{\mu_1}...\hat{\xi}^{\mu_q}=\hat{\xi}^{\mu_1}...\hat{\xi}^{\mu_q}\hat{\tilde{\omega}}_{\mu_1 ... \mu_q},\\
\hat{\theta}&=\hat{\theta}_{\mu_1 ... \mu_p}\hat{\xi}^{\mu_1}...\hat{\xi}^{\mu_p}=\hat{\xi}^{\mu_1}...\hat{\xi}^{\mu_p}\hat{\tilde{\theta}}_{\mu_1 ... \mu_p},
\end{split}\end{equation}
where $\hat{\omega}_{\mu_1 ... \mu_q}$, $\hat{\tilde{\omega}}_{\mu_1 ... \mu_q}$, $\hat{\theta}_{\mu_1 ... \mu_p}$ and $\hat{\tilde{\theta}}_{\mu_1 ... \mu_p}$ $\in\mathcal{\hat{A}}$. There is a relation between $\hat{\omega}_{\mu_1 ... \mu_q}$ and $\hat{\tilde{\omega}}_{\mu_1 ... \mu_q}$ via \eqref{lambda} (same for $\hat{\theta}_{\mu_1 ... \mu_p}$ and $\hat{\tilde{\theta}}_{\mu_1 ... \mu_p}$ ). Of course if we multiply $\hat{\omega}$ with $\hat{\theta}$ it is easy to see that $\hat{\omega}\hat{\theta}\neq\hat{\theta}\hat{\omega}$, but because of bicovarince we have 
\begin{equation}\begin{split}
\hat{\omega}\hat{\theta}&=\hat{\alpha}_{\mu_1 ... \mu_{q+p}}\hat{\xi}^{\mu_1}...\hat{\xi}^{\mu_{q+p}}=\hat{\xi}^{\mu_1}...\hat{\xi}^{\mu_{q+p}}\hat{\tilde{\alpha}}_{\mu_1 ... \mu_{q+p}}\in\hat{\Omega}^{q+p}\\
\hat{\theta}\hat{\omega}&=\hat{\beta}_{\mu_1 ... \mu_{q+p}}\hat{\xi}^{\mu_1}...\hat{\xi}^{\mu_{q+p}}=\hat{\xi}^{\mu_1}...\hat{\xi}^{\mu_{q+p}}\hat{\tilde{\beta}}_{\mu_1 ... \mu_{q+p}}\in\hat{\Omega}^{q+p}
\end{split}\end{equation}
where $\hat{\alpha}_{\mu_1 ... \mu_{q+p}}$, $\hat{\tilde{\alpha}}_{\mu_1 ... \mu_{q+p}}$, $\hat{\beta}_{\mu_1 ... \mu_{q+p}}$ and $\hat{\tilde{\beta}}_{\mu_1 ... \mu_{q+p}}$ $\in\hat{\mathcal{A}}$ and they are interrelated with $\hat{\omega}_{\mu_1 ... \mu_q}$ and $\hat{\theta}_{\mu_1 ... \mu_p}$ by using \eqref{lambda}. Bicovariance leads to condition \eqref{forme} which enables us to find \eqref{lambda} and define all the higher forms in a consistent way (as illustrated above).

In formulating differential calculus it is important to know all the commutation rules for (and between) the elements of $\hat{\mathcal{A}}$ and $\hat{\Omega}^p$. The commutation rule between one form $\hat{\xi}_{\mu}$ and an arbitrary element of $\hat{\mathcal{A}}$ is determined by $\Lambda_{\mu\nu}$ \eqref{lambda}. The commutation rule between NC coordinate $\hat{x}_{\mu}$ and an arbitrary element of $\hat{\mathcal{A}}$ is determined by $O_{\mu\nu}$ (see Appendix A)
\begin{equation} \label{xffx}
\hat{x}_{\mu}\hat{f}=(O_{\mu\alpha}\blacktriangleright\hat{f}) \hat{x}^{\alpha},  \quad \hat{f}\hat{x}_{\mu}=\hat{x}^{\alpha}([O^{-1}]_{\mu\alpha}\blacktriangleright\hat{f}).
\end{equation}
We point out that
\begin{equation}
O_{\mu\nu}=\left(\text{e}^{\EuScript{C}}\right)_{\mu\nu}, \qquad \EuScript{C}_{\mu\nu}=iC_{\mu\alpha\nu}(\partial^{W})^{\alpha},
\end{equation}
and similarly $\Lambda_{\mu\nu}$ is given by\footnote{For a different derivation of $\Lambda_{\mu\nu}$ using realization and coproduct see \cite{linreal}.} 
\begin{equation}\label{lambdaexp}
\Lambda_{\mu\nu}=\left(\text{e}^{\EuScript{K}}\right)_{\mu\nu}, \qquad \EuScript{K}_{\mu\nu}=iK_{\mu\alpha\nu}(\partial^{W})^{\alpha},
\end{equation}
where  $\partial^W$ is the derivative corresponding to the Weyl  ordering \cite{Meljanac-3} (see also Appendix A and B):
 \begin{equation}
 [\partial^W_\mu,\hat x_\nu]=\eta_{\mu\nu} \frac{ia\partial^W}{\text e^{ia\partial^W}-1}
 +\frac{ia_\nu\partial^W_\mu}{ia\partial^W} \left(1-\frac{ia\partial^W}{\text e^{ia\partial^W}-1}\right),
 \end{equation}
 where we used $a\partial^W \equiv a^\alpha \partial^W _{\alpha}$.
We also have 
\begin{align} \label{Lambdaxcomm}
[\Lambda_{\mu\alpha},\hat x_\nu]&=iK_{\mu\nu}{}^\beta\Lambda_{\beta\alpha}, \\
\label{Oxcomm}
[O_{\mu\alpha},\hat x_\nu]&=iC_{\mu\nu}{}^\beta O_{\beta\alpha}.
\end{align}
In $\kappa$-Minkowski space it is useful to introduce the shift operator $Z$, defined by 
\begin{equation}\begin{split}
[Z,\hat x_\mu] &= ia_\mu Z, \quad [Z,\partial_\mu]=0, \\
[Z,\hat\xi_\mu]&=0, \quad [Z,q_\mu]=0, 
\end{split}\end{equation}
where 
\begin{equation}\label{Za}
Z=\text e^{ia\partial^W}.
\end{equation}
The expression for $O_{\mu\nu}$ is:
\begin{equation}
O_{\mu\nu} = \eta_{\mu\nu}Z^{-1} + ia_\mu\partial^W_\nu \frac{1-Z^{-1}}{\ln Z}.
\end{equation}
Explicit expressions for $\Lambda_{\mu\nu}$ (and its inverse), $\partial_\mu^W$ and $Z$ for $\mathcal S_1$, $\mathcal S_2$, $\mathcal S_3$ and $\mathcal C_4$ are:
\begin{itemize}
\item $\mathcal{S}_1$:
\begin{equation}
\Lambda_{\mu\nu}= \eta_{\mu\nu}Z^{-1}, \quad [\Lambda^{-1}]_{\mu\nu}=\eta_{\mu\nu}Z,
\end{equation}
\begin{equation}
\partial^W_\mu = \partial^R_\mu \frac{\ln Z}{Z-1}, \quad Z=\text{e}^{ia\partial^W}=1+ia\partial^R
\end{equation}
\item $\mathcal{S}_2$:
\begin{equation}
\Lambda_{\mu\nu}= \eta_{\mu\nu} + ia_\mu \partial^W_\nu \frac{Z-1}{\ln Z}
=\eta_{\mu\nu} + ia_\mu \partial^{L}_{\nu} Z, \quad [\Lambda^{-1}]_{\mu\nu}=\eta_{\mu\nu}-ia_{\mu}\partial^{L}_{\alpha},
\end{equation}
\begin{equation}
\partial_\mu^W = \partial^L_\mu \frac{\ln Z}{1-Z^{-1}}, \quad Z=\frac{1}{1-ia\partial^L}
\end{equation}
\item $\mathcal{S}_3$:
\begin{equation}
\Lambda_{\mu\nu}= \left(\eta_{\mu\nu}-ia_\nu \partial^W_\mu \frac{1-Z^{-1}}{\ln Z} \right)Z^{-1}
, \quad [\Lambda^{-1}]_{\mu\nu}=\eta_{\mu\nu}Z+ia_{\nu}\partial^{W}_{\mu}\frac{Z(Z-1)}{\ln{Z}}
\end{equation}
\begin{equation}\begin{split}
\partial_\mu^W = \left( \partial^{MS}_\mu - ia_\mu\frac{Z-\sqrt{Z^2-a^2(\partial^{MS})^2}}{a^2} \right)&\frac{\ln Z}{Z-1}, \quad  Z=\sqrt{(1+ia\partial^{MS})^2+a^2(\partial^{MS})^2}\\
\end{split}\end{equation}
\item $\mathcal{C}_4 \ \ (a^2\overset{!}{=}0)$:
\begin{equation}\begin{split}
\Lambda_{\mu\nu}&=\eta_{\mu\nu}+ia_\mu \partial^W_\nu \frac{Z-1}{\ln Z}
+ia_\nu \partial^W_\mu \frac{Z^{-1}-1}{\ln Z}
+\frac{a_\mu a_\nu}2 (\partial^W)^2 \left(\frac{Z^{1/2}-Z^{-1/2}}{\ln Z} \right)^2\\
&=\eta_{\mu\nu}-ia_{\nu}D_{\mu}+ia_{\mu}(D_{\nu}-\frac{i}{2}a_{\nu}D^2)Z\\
[\Lambda^{-1}]_{\mu\nu}&=\eta_{\mu\nu}-ia_{\mu}D_{\nu}+ia_{\nu}(D_{\mu}-\frac{i}{2}a_{\mu}D^2)Z
\end{split}\end{equation}
\begin{equation}
\partial_\mu^W = \left(D_\mu -\frac{ia_\mu}2 D^2 \right)\frac{\ln Z}{1-Z^{-1}}, \quad Z=\frac{1}{1-iaD}
\end{equation}
\end{itemize}
where $\partial^R_\mu$, $\partial^L_\mu$, $\partial^{MS}_\mu$  and $D_\mu$ correspond to $\mathcal S_1$, $\mathcal S_2$, $\mathcal S_3$ and $\mathcal C_4$ respectively (for the derivation see Appendix B).

The commutation rule between NC coordinate $\hat x_\mu$ and an arbitrary element $\hat\chi=\hat\xi_{i_1}...\hat\xi_{i_p}$, $p\le n$ is
\begin{equation}
[\hat\chi,\hat x_\mu]=\hat{\xi}^{\beta}\left(iK_{\alpha\mu\beta}q^\alpha\blacktriangleright\hat\chi\right).
\end{equation}

From $\hat x_\mu=x_\mu-iK_{\alpha\mu}{}^\rho(x_\rho\partial^\alpha+\xi_\rho q^\alpha)=x_\mu-iK_{\alpha\mu}{}^\rho L_\rho{}^\alpha$, where $L_\rho{}^\alpha$ generates the $\mathfrak{gl}(n)$  algebra (see eq.~(74a) in Section VI), it follows
\begin{equation}
[q_\mu,\hat x_\nu]=-iq^\beta K_{\beta\nu\mu}
\end{equation}
and we find
\begin{equation}
q_{\mu}\hat{f}=(q_{\mu}\blacktriangleright\hat{f})+(\Lambda^{-1}_{\alpha\mu}\blacktriangleright\hat{f})q^{\alpha}
\end{equation}
from which we get the Leibniz rule for $q_{\mu}$ and extract the following coproduct
\begin{equation}\label{deltaq}
\Delta q_{\mu}=q_{\mu}\otimes 1+(-)^{N_1}[\Lambda^{-1}]_{\alpha\mu}\otimes q^{\alpha},
\end{equation}
where $[N_1, \hat x_\mu]=[N_1, \partial_\mu]=0$, $[N_1,\hat\xi_\mu]=\hat\xi_\mu$ and $[N_1,q_\mu]=-q_\mu$.
Note that $\partial_{\mu}$, $q_{\mu}$, $\hat{\eta}$ and $N_1$ generate a super-Hopf algebra. From the undeformed Leibniz rule for $\hat{\eta}$ it follows that $\Delta\hat\eta$ is undeformed, i.e. $\Delta \hat\eta = \Delta_0 \hat\eta = \hat\eta\otimes1+(-)^{N_1}\otimes\hat\eta$ and $\Delta N_1 = N_1 \otimes 1 + 1 \otimes N_1$ (see section VI).


\section{Related class of Drinfeld twists}

Here we present Drinfeld twists leading to equations \eqref{real} and \eqref{crealizacije} which generate super Hopf algebra structure (sections IV and VI). These twists can be obtained using a method that has been developed for constructing a Drinfeld twist from linear realizations \eqref{real} \cite{linreal}
\begin{equation}
\mathcal F= \exp\left( \mathcal K_{\beta\alpha} \otimes L^{\alpha\beta} \right)
\label{twistF}
\end{equation}
where 
$L_{\mu\nu}$ generate $\mathfrak{gl}(n)$ algebra, see equations \eqref{algebraL} and $\mathcal K_{\mu\nu}=iK_{\mu\alpha\nu}(\partial^W)^\alpha$, see equation \eqref{lambdaexp}.
They lead to  the super-Hopf algebra structures defined by $\Delta \partial_\mu$, $\Delta q_\mu$, $\Delta L_{\mu\nu}$, where the coproduct, antipode and counit are:
\begin{equation}\begin{split}
&\Delta g=\mathcal{F}\Delta_{0}g\mathcal{F}^{-1},  \\
&S(g)=\chi \ S_{0}(g)\ \chi^{-1}, \quad \chi= m\left[\left(1 \otimes S_{0}\right)\mathcal{F}\right], \\
&\epsilon(g)=0, 
\end{split}\end{equation}
where $\Delta_{0}g=g\otimes 1+1\otimes g$ is the undeformed coproduct for $\partial_\mu$ and $L_{\mu\nu}$,  $\Delta_0 q_\mu = q_\mu \otimes 1 + (-)^{N_1} \otimes q_\mu$ and $S_0(g)=-g$ is the undeformed antipode for $\partial_\mu$ and $L_{\mu\nu}$ and $S_0(q_\mu)=-(-)^{N_1}q_\mu=q_\mu(-)^{N_1}$. Drinfeld twists \eqref{twistF} are new and lead to deformed super Hopf algebra in section VI.

For $\mathcal C_1$, $\mathcal C_2$, $\mathcal C_3$ and $\mathcal C_4$, twists are:
\begin{align}
\label{twistC1}
&\mathcal F_{\mathcal C_1} = \exp \left\{
\left( \eta_{\alpha\beta} - c\frac{a_\alpha a_\beta}{a^2} \right) \ln Z
\otimes L^{\alpha\beta} \right\}
=\exp \left\{
\ln Z \otimes \left(D-c\frac{a_\alpha a_\beta}{a^2}L^{\alpha\beta} \right)
\right\} 
, \\
\label{twistC2}
&\mathcal F_{\mathcal C_2} = \exp \left\{
\left[
c\left( \eta_{\alpha\beta} - \frac{a_\alpha a_\beta}{a^2} \right) \ln Z
+ia_\beta \partial^W_\alpha
\right]
\otimes L^{\alpha\beta} \right\}, \\
\label{twistC3}
&\mathcal F_{\mathcal C_3} = \exp \left\{
\left[
\left( \eta_{\alpha\beta} - c\frac{a_\alpha a_\beta}{a^2} \right) \ln Z
+ (1+c)ia_\alpha \partial^W_\beta
\right]
\otimes L^{\alpha\beta} \right\}, \\
\label{twistC4}
&\mathcal F_{\mathcal C_4} =
\exp \left\{
\left(ia_\alpha \partial^W_\beta-ia_\beta \partial^W_\alpha \right) \otimes L^{\alpha\beta}
\right\} = 
\exp \left\{
ia_\alpha \partial^W_\beta \otimes M^{\alpha\beta}
\right\}
\end{align}
where $M_{\mu\nu}=L_{\mu\nu}-L_{\nu\mu}$. Note that in \cite{EPJC} the extended twist was used to obtain the differential algebra $\mathcal{D}_{1}^{c=1}=\mathcal{D}_{2}^{c=1}$.

Starting from the twist operator, the realization can be obtained using
\begin{equation}
\hat x_\mu = m\left[\mathcal F^{-1} (\triangleright \otimes 1)(x_\mu \otimes 1) \right] =
 x_\mu - iK_{\beta\mu\alpha}L^{\alpha\beta}
\end{equation}
Using twists \eqref{twistC1}, \eqref{twistC2}, \eqref{twistC3} and \eqref{twistC4} yields realizations $\mathcal C_1$, $\mathcal C_2$, $\mathcal C_3$ and $\mathcal C_4$ respectively, which satisfy $\kappa$-Minkowski algebra.

\subsection{Twist leading to $\kappa$-Poincar\'e Hopf algebra}
 Here we point out that only the case $\mathcal C_4$ is $\kappa$-Poincar\'e covariant and that the corresponding twist operator can be written in a covariant form \cite{universal}
\begin{equation}
\mathcal F_{\mathcal C_4}=\text{exp}\left\{a^\alpha P^\beta\frac{\text{ln}(1+a\cdot P)}{a\cdot P}\otimes M_{\alpha\beta}\right\}, \quad a^2=0,
\label{naturaltwist}
\end{equation}
which is expressed in terms of Poincar\'e generators only and satisfies the cocycle condition.  

We point out that the twist in eq. \eqref{naturaltwist} is a unique covariant Drinfeld twist, expressed in terms of Poincar\'e generators, compatible with $\kappa$-Minkowski space \eqref{kappaminkowski} and $\kappa$-Poincar\'e (super) Hopf algebra:
\begin{equation}\begin{split} \label{kphalgebra}
\Delta P_\mu &=  \Delta_0 P_\mu + \left[
P_\mu a^\alpha - a_\mu
\left( P^\alpha + \frac{1}{2}a^\alpha P^2 \right)Z
\right]\otimes P_\alpha
\\
\Delta Q_\mu &= \Delta_0 Q_\mu + (-)^{N_1} \left[ 
P_\mu a^\alpha - a_\mu
\left( P^\alpha + \frac{1}{2}a^\alpha P^2 \right)Z
\right]\otimes Q_\alpha 
\\
\Delta M_{\mu\nu} &= \Delta_0 M_{\mu\nu} +
(\delta^\alpha_\mu a_\nu-\delta^\alpha_\nu a_\mu)\left(
P^\beta+\frac{1}{2}a^\beta P^2
\right)Z\otimes M_{\alpha\beta}
\\
S(P_\mu) &= \left[-P_\mu -a_\mu \left(P_\alpha + \frac{1}{2}a_\alpha P^2 \right) P^\alpha \right]Z \\
S(Q_\mu) &=(-)^{N_1} \left[-Q_\mu -a_\mu \left(P_\alpha + \frac{1}{2}a_\alpha P^2 \right) Q^\alpha \right]Z \\
S(M_{\mu\nu}) &= 
-M_{\mu\nu} +(-a_\mu \delta^\beta_\nu+a_\nu \delta^\beta_\mu) \left(P^\alpha + \frac{1}{2}a^\alpha P^2 \right)M_{\alpha\beta} 
\end{split}\end{equation}

and the algebra is given by
\begin{equation}\begin{split}
[M_{\mu\nu},M_{\lambda\rho}]=-i(\eta_{\nu\lambda}M_{\mu\rho}&-\eta_{\mu\lambda}M_{\nu\rho}-\eta_{\nu\rho}M_{\mu\lambda}+\eta_{\mu\rho}M_{\nu\lambda})\\
[M_{\mu\nu},P_{\lambda}]&=\eta_{\nu\lambda}P_{\mu}-\eta_{\mu\lambda}P_{\nu}\\
[M_{\mu\nu},Q_{\lambda}]&=\eta_{\nu\lambda}Q_{\mu}-\eta_{\mu\lambda}Q_{\nu}\\
[P_{\mu}, Q_{\nu}]=&[P_{\mu},P_{\nu}]=\left\{Q_{\mu},Q_{\nu}\right\}=0.
\end{split}\end{equation}




\section{Super-algebra $\mathcal{L}$ and deformed super-Hopf algebras}

The differential calculi that we developed so far is bicovariant. Bicovariance condition states that one-forms $\hat{\xi}_{\mu}$ are simultaneously left and right covariant \cite{19, woro}. The sufficient condition for bicovariance is given by
\begin{equation}\label{com}
[\hat{\xi}_{\mu},\hat{x}_{\nu}]=iK_{\mu\nu}{}^\alpha \hat{\xi}_{\alpha},
\end{equation}
 that is that the commutator \eqref{com} is closed in one-forms (and the differential calculus is of classical dimension). Of course conditions \eqref{conditionK} and \eqref{consist} naturally hold. Covariance of differential calculus under a certain symmetry algebra $\mathcal{G}\subset\hat{\mathcal{SH}}$ generated by $g_i \in\mathcal{G}$ is defined in the following way
\begin{equation}
g_i\blacktriangleright(\hat{x}_{\mu}\hat{\xi}_{\nu})=m\left(\Delta g_{i}(\blacktriangleright\otimes \blacktriangleright)(\hat{x}_{\mu}\otimes\hat{\xi}_{\nu})\right), \quad \text{and} \quad g_i\blacktriangleright(\hat{\xi}_{\nu}\hat{x}_{\mu})=m\left(\Delta g_{i}(\blacktriangleright\otimes \blacktriangleright)(\hat{\xi}_{\nu}\otimes\hat{x}_{\mu})\right).
\end{equation}

The question that remains is what are the symmetries of our differential calculi. To answer this question we will consider an algebra $\mathcal{L}$ which is generated by $L_{\mu\nu}$, $\partial_{\mu}$, $q_{\mu}$, $\eta$, $N_0$ and $N_1$ and has a super-Hopf algebra structure. 

\subsection{Undeformed super-Hopf algebra structure of $\mathcal{L}$}

The super-algebra $\mathcal{L}$ is generated by $\mathfrak{gl}(n)$ generators $L_{\mu\nu}$, translations $\partial_{\mu}$ and $q_{\mu}$, generators $\eta$, $N_{0}$ and $N_{1}$ satisfying 
\begin{subequations}\label{algebraL}\begin{equation}
[L_{\mu\nu},L_{\alpha\beta}]=\eta_{\nu\alpha}L_{\mu\beta}-\eta_{\mu\beta}L_{\alpha\nu},
\end{equation}
\begin{equation}
[L_{\mu\nu},\partial_{\rho}]=-\eta_{\mu\rho}\partial_{\nu}, \quad [L_{\mu\nu},q_{\rho}]=-\eta_{\mu\rho}q_{\nu}, \quad [L_{\mu\nu},\eta]=0,
\end{equation}
\begin{equation}
[\partial_{\mu},\partial_{\nu}]=\left\{q_{\mu},q_{\nu}\right\}=[\partial_{\mu}, q_{\nu}]=0,
\end{equation}
\begin{equation}
\eta^2=0, \quad \left\{\eta,q_{\mu}\right\}=\partial_{\mu}, \quad [\eta, \partial_{\mu}]=0, 
\end{equation}
\begin{equation}
[N_0, L_{\mu\nu}]=[N_1, L_{\mu\nu}]=[N_{0}, N_1]=0,
\end{equation}
\begin{equation}
[N_0, \partial_{\mu}]=-\partial_{\mu}, \quad [N_{0}, q_{\mu}]=[N_{1}, \partial_{\mu}]=0,
\end{equation}
\begin{equation}
[N_{1},q_{\mu}]=-q_{\mu}, \quad [N_{0}, \eta]=-\eta, \quad [N_1, \eta]=\eta.
\end{equation}
\end{subequations}
There exists an undeformed super-Hopf algebra structure defined by the coalgebra structure
\begin{equation}
\Delta_{0}g=g\otimes 1 +1\otimes g,
\end{equation}
where $g\in\left\{L_{\mu\nu},\partial_{\mu}, N_{0}, N_{1}\right\}$, with grade $|g|=0$ and coproducts for $q_{\mu}$ and $\eta$ are given by
\begin{equation}
\Delta_{0}q_{\mu}=q_{\mu}\otimes 1+(-)^{N_1}\otimes q_{\mu},
\end{equation}
\begin{equation}
\Delta_{0}\eta=\eta\otimes 1+(-)^{N_{1}}\otimes \eta,
\end{equation}
with grade $|q|=1$ and $|\eta|=1$.

The antipode $S_{0}$ is defined by
\begin{equation}
S_{0}(g)=-g
\end{equation}
for $g\in\left\{L_{\mu\nu},\partial_{\mu}, N_{0}, N_{1}\right\}$ and 
\begin{equation}
S_{0}(q_{\mu})=-(-)^{N_1}q_{\mu}=q_{\mu}(-)^{N_1}, \quad S_{0}(\eta)=-(-)^{N_1}\eta=\eta(-)^{N_1}.
\end{equation}
The counit $\epsilon_{0}$ is defined by
\begin{equation}
\epsilon_0(g)=0,
\end{equation}
for all generators, and $\epsilon_0(1)=1$.

\subsection{Deformed super-Hopf algebra structure of $\mathcal{L}$}

Using Drinfeld twists from section V we construct deformed super-Hopf algebras which are defined by the coalgebra structure $\Delta$
\begin{equation}
\Delta L_{\mu\nu}=L_{\mu\nu}\otimes1+\left(\Lambda^{-1}_{\beta\gamma} \frac{\partial \Lambda^\gamma{}_\alpha}{\partial (\partial^{\mu})}\partial_{\nu}+\Lambda^{-1}_{\beta\nu}\Lambda_{\mu \alpha}\right)\otimes L^{\alpha\beta},
\end{equation}
\begin{equation}
\Delta\partial_{\mu}=\partial_{\mu}\otimes 1+\Lambda^{-1}_{\alpha\mu}\otimes\partial^{\alpha},
\end{equation}
\begin{equation}
\Delta q_{\mu}=q_{\mu}\otimes 1+(-)^{N_1}\Lambda^{-1}_{\alpha\mu}\otimes q^{\alpha},
\end{equation}
\begin{equation}
\Delta \eta=\eta\otimes 1+(-)^{N_1}\otimes \eta,
\end{equation}
\begin{equation}
\Delta N_0=\Delta_0 N_0
+ \Lambda^{-1}_{\beta\gamma} \frac{\partial \Lambda^\gamma{}_\alpha}{\partial (\partial_\mu)}\partial_\mu \otimes L^{\alpha\beta}
\end{equation}
\begin{equation}
\Delta N_1=N_1 \otimes 1 + 1 \otimes N_1=\Delta_0 N_1
\end{equation}
where $\Lambda_{\mu\nu}=(\text{e}^{\EuScript{K}})_{\mu\nu}$ and $\EuScript{K}_{\mu\nu}=iK_{\mu\alpha\nu}(\partial^{W})^{\alpha}$. Antipode $S$ for all generators is obtained by using $S(\partial_\mu^W)=-\partial_\mu^W$, $S(\EuScript K_{\mu\nu})=-\EuScript K_{\mu\nu}$, $S(\Lambda_{\mu\nu})=\Lambda^{-1}_{\mu\nu}$, $S(\Lambda^{-1}_{\mu\nu})=\Lambda_{\mu\nu}$, and $S(\partial_\mu)=-\partial_\alpha\Lambda^\alpha{}_\mu$.
The counit $\epsilon$ is unchanged, i.e. $\epsilon=\epsilon_0$. This defines the deformed super-Hopf algebra structure of $\mathcal{L}$.

Since $\mathcal{L}\subset\hat{\mathcal{SH}}$ for the action $\blacktriangleright$ (see \eqref{crnodjelovanje}) we have
\begin{equation}\begin{aligned}
L_{\mu\nu}\blacktriangleright\hat{x}_{\lambda}&=\eta_{\nu\lambda}\hat{x}_{\mu}, &
L_{\mu\nu}\blacktriangleright\hat{\xi}_{\lambda}&=\eta_{\nu\lambda}\hat{\xi}_{\mu}, &
\eta \blacktriangleright \hat x_\mu &=\hat\xi_\mu, & 
\eta \blacktriangleright \hat \xi_\mu &= 0, 
\\
N_0\blacktriangleright\hat{x}_{\mu}&=\hat{x}_{\mu}, & 
N_0\blacktriangleright\hat{\xi}_{\mu}&=0, & 
N_1\blacktriangleright\hat{x}_{\mu}&=0, &  
N_1\blacktriangleright\hat{\xi}_{\mu}&=\hat{\xi}_{\mu}, 
\\
\partial_{\mu}\blacktriangleright\hat{x}_{\nu}&=\eta_{\mu\nu}, & 
\partial_{\mu}\blacktriangleright\hat{\xi}_{\nu}&=0, &
q_{\mu}\blacktriangleright\hat{x}_{\nu}&=0, & 
q_{\mu}\blacktriangleright\hat{\xi}_{\nu}&=\eta_{\mu\nu} 
\end{aligned}\end{equation}
and $g\blacktriangleright1=0$, $\forall g \in \mathcal L$.

In order to calculate the action of the generators of $\mathcal L$ on an arbitrary element of $\hat{\mathcal{SA}}$, we use commutation relations $[L_{\mu\nu},\hat x_\lambda]=\eta_{\nu\lambda}(\hat x_\mu + iK_{\beta\mu\alpha}L^{\alpha\beta})-iK_{\beta\lambda\nu}L_\mu{}^\beta+iK_{\mu\lambda\alpha}L^\alpha{}_\nu$ and $[L_{\mu\nu}, \hat\xi_\lambda]=\eta_{\lambda\nu}\hat\xi_\mu$, which are obtained using $L_{\mu\nu} = x_\mu\partial_\nu + \xi_\mu q_\nu$, $\hat x_\mu = x_\mu - iK_{\beta\mu\alpha}L^{\alpha\beta}$ and $\hat\xi_\mu=\xi_\mu$.

The covariance property under the super-Hopf algebra $\mathcal L$ holds:
\begin{equation}
g \blacktriangleright (\hat f \hat g) = 
m \left( \Delta g  (\hat f \otimes \hat g) \right), 
\quad g \in \mathcal L,  \quad \hat f, \hat g \in \hat{\mathcal{SA}}
\end{equation}
For example, one can easily check
\begin{equation}
L_{\mu\nu}\blacktriangleright(\hat \xi_\alpha \hat x_\beta)=
\eta_{\nu\alpha}\hat\xi_\mu \hat x_\beta 
+ \eta_{\nu\beta}\hat\xi_\alpha \hat x_\mu
=m\left(\Delta L_{\mu\nu}\blacktriangleright(\hat \xi_\alpha \otimes \hat x_\beta)\right),
\end{equation}
\begin{equation}
L_{\mu\nu}\blacktriangleright(\hat x_\beta \hat \xi_\alpha)=
\eta_{\nu\alpha}\left(\hat x_\beta \hat \xi_\mu + iK_{\mu\beta}{}^\gamma \hat \xi_\gamma \right)+
\eta_{\nu\beta}\left(\hat x_\mu \hat \xi_\alpha + iK_{\alpha\mu}{}^\gamma \hat \xi_\gamma \right)-
iK_{\alpha\beta\nu} \hat \xi_\mu
=m\left(\Delta L_{\mu\nu}\blacktriangleright(\hat x_\beta \otimes \hat \xi_\alpha)\right),
\end{equation}
and
\begin{equation}
L_{\mu\nu}\blacktriangleright[\hat \xi_\alpha, \hat x_\beta]=
iK_{\alpha\beta}{}^\gamma \hat \xi_\gamma
=m\left(\Delta L_{\mu\nu}\blacktriangleright(\hat \xi_\alpha \otimes \hat x_\beta - \hat x_\beta \otimes \hat \xi_\alpha)\right).
\end{equation}

The deformed super-Hopf algebra acting on $\hat{x}_{\mu}\otimes 1$ and $\hat{\xi}_{\mu}\otimes 1$, i.e. using $g\hat f=m\left(\Delta g(\blacktriangleright\otimes 1)(\hat f\otimes 1)\right)$, $\forall g\in\mathcal L$ and $\hat f \in\hat{\mathcal{SA}}$, leads to
\begin{equation}\label{93}
[L_{\rho\sigma}, \hat{x}_{\nu}]=\eta_{\sigma\nu}\hat{x}_{\rho}+i\eta_{\sigma\nu}K_{\mu\rho\alpha}L^{\alpha\mu}-iK_{\mu\nu\sigma}L_{\rho}^{\ \mu}+iK_{\rho\nu\alpha}L^{\alpha}_{\ \sigma}
\end{equation}
\begin{equation}
[L_{\mu\nu},\hat{\xi}_{\lambda}]=\eta_{\nu\lambda}\hat{\xi}_{\mu},
\end{equation}
\begin{equation}
[\partial_{\mu}, \hat{x}_{\nu}]=\eta_{\mu\nu}-iK_{\beta\nu\mu}\partial^{\beta}, \quad [\partial_{\mu}, \hat{\xi}_{\nu}]=0,
\end{equation}
\begin{equation}
[q_{\mu}, \hat{x}_{\nu}]=-iK_{\beta\nu\mu}q^{\beta}, \quad \left\{q_{\mu}, \hat{\xi}_{\nu}\right\}=\eta_{\mu\nu},
\end{equation}
\begin{equation}
[\eta, \hat{x}_{\mu}]=\hat{\xi}_{\mu}, \quad \left\{\eta,\hat{\xi}_{\mu}\right\}=0,
\end{equation}
which also leads to
\begin{equation}
[\hat{x}_{\mu},\hat{x}_{\nu}]=a_{\mu}\hat{x}_{\nu}-a_{\nu}\hat{x}_{\mu},
\end{equation}
\begin{equation}\label{99}
\left\{\hat{\xi}_{\mu},\hat{\xi}_{\nu}\right\}=0,\quad [\hat{\xi}_{\mu},\hat{x}_{\nu}]=iK^{\ \ \ \alpha}_{\mu\nu}\hat{\xi}_{\alpha}.
\end{equation}

The realization for $\hat{x}_{\mu}$, $\hat{\xi}_{\mu}$, $L_{\mu\nu}$, $N_0$, $N_1$ and $\eta$ in terms of the super-Heisenberg algebra $\mathcal{SH}$ follows from \eqref{93}-\eqref{99} and  is given by
\begin{subequations}
\begin{equation}
\hat{x}_{\mu}=x_{\mu}-iK_{\beta\mu\alpha}L^{\alpha\beta},
\end{equation}
\begin{equation}
\hat{\xi}_{\mu}=\xi_{\mu},
\end{equation}
\begin{equation}
L_{\mu\nu}=x_{\mu}\partial_{\nu}+\xi_{\mu}q_{\nu},
\end{equation}
\begin{equation}
N_0=x_{\alpha}\partial^{\alpha},
\end{equation}
\begin{equation}
N_1=\xi_{\alpha}q^{\alpha},
\end{equation}
\begin{equation}
\eta=\xi_{\alpha}\partial^{\alpha}.
\end{equation}
\end{subequations}
This proves the consistency of our approach.

We point out that generators $l_\mu = \hat x_\mu - x_\mu = -i K_{\beta\mu\alpha}L^{\alpha\beta}$ close $\kappa$-Minkowski algebra (when $a_\mu \rightarrow 0$ then $l_\mu \rightarrow 0$). Since $l_\mu \in \mathcal L$ and $\Delta_0 l_\mu = l_\mu \otimes 1 + 1 \otimes l_\mu$, we can apply twist $\mathcal F$ from equation \eqref{twistF}
\begin{equation}
\Delta l_\mu = \mathcal F \Delta_0 l_\mu \mathcal F^{-1}.
\end{equation}
From general structure of twist $\mathcal F = \exp\left(\mathcal K_{\beta\alpha}\otimes L^{\alpha\beta}\right)=\exp\left(-\partial^W_\alpha\otimes l^\alpha\right)$ it is clear that $\Delta l_\mu$ is closed in $l_\alpha$ and $\partial_\alpha$. 
This is the main feature of all solutions in the present paper.

\subsection{Symmetries of differential algebras $\mathcal{S}_{1}$, $\mathcal{S}_{2}$, $\mathcal{S}_{3}$ and $\mathcal{C}_{4}$}

Our differential algebras (obtained in section II) are covariant under certain $\kappa$-deformation of the $\mathfrak{igl}(n)$ algebra, but in the special case of $\mathcal{S}_1$ we have  Poincar\'e -Weyl, and in the case of $\mathcal{C}_4$ $\kappa$- Poincar\'e  covariance.

For $\mathcal{S}_{1}$ defined by $K_{\mu\nu\alpha}=-\eta_{\mu\alpha}a_{\nu}$ we get $\hat x_\mu=x_\mu +ia_\mu D$ and we have 
\begin{equation}
[L_{\rho\sigma}, \hat{x}_{\nu}]=\eta_{\sigma\nu}\hat{x}_{\rho}-i\eta_{\sigma\nu}a_{\rho}D,
\end{equation}
where $D\equiv L_\alpha{}^\alpha=N_0+N_1$. If we define the Lorentz($SO(1,n-1)$) generators by $M_{\mu\nu}=L_{\mu\nu}-L_{\nu\mu}$ we have 
\begin{equation}\label{poincare-weyl}
[M_{\rho\sigma}, \hat{x}_{\nu}]=\eta_{\sigma\nu}\hat{x}_{\rho}-\eta_{\rho\nu}\hat{x}_{\sigma}-i(\eta_{\sigma\nu}a_{\rho}-\eta_{\rho\nu}a_{\sigma})D, \quad [D, \hat{x}_{\lambda}]=\hat{x}_{\lambda}-ia_{\lambda}D.
\end{equation}
Note that eq. \eqref{poincare-weyl} includes the dilatation operator $D$. This implies the  Poincar\'e -Weyl algebra, which is the underlying symmetry of the differential algebra $\mathcal{S}_{1}$ and is compatible with \cite{21, 22}.

For $\mathcal{S}_{2}$ defined by $K_{\mu\nu\alpha}=\eta_{\nu\alpha}a_{\mu}$ we get $\hat x_\mu = x_\mu-ia^\alpha L_{\mu\alpha}$ and we have 
\begin{equation}
[L_{\rho\sigma}, \hat{x}_{\nu}]=\eta_{\sigma\nu}\hat{x}_{\rho}+ia_{\rho}L_{\nu\sigma},
\end{equation}
which illustrates that the underlying symmetry of the differential algebra $\mathcal{S}_{2}$ is described by the $\kappa$-deformed $\mathfrak{igl}(n)$-algebra and is compatible with \cite{beggs}.

For $\mathcal{S}_{3}$ defined by $K_{\mu\nu\alpha}=-\eta_{\mu\nu}a_{\alpha}-\eta_{\mu\alpha}a_{\nu}$ we get $\hat x_\mu = x_\mu + ia^\alpha L_{\alpha\mu} + ia_\mu D$ and we have 
\begin{equation}
[L_{\rho\sigma}, \hat{x}_{\nu}]=\eta_{\sigma\nu}\hat{x}_{\rho}-i\eta_{\sigma\nu}a^{\alpha}L_{\alpha\rho}+ia_{\sigma}L_{\rho\nu}-i\eta_{\rho\nu}a_{\alpha}L^{\alpha}_{\ \sigma}-i\eta_{\sigma\nu}a_{\rho}D,
\end{equation}
which illustrates that the underlying symmetry of the differential algebra $\mathcal{S}_{3}$ is described by the $\kappa$-deformed $\mathfrak{igl}(n)$-algebra and is compatible with the Magueijo-Smolin realization \cite{kovacevic-meljanac}.

For $\mathcal{C}_{4}$ defined by $K_{\mu\nu\alpha}=-\eta_{\mu\nu}a_{\alpha}+\eta_{\nu\alpha}a_{\mu}, \ \ a^2=0$, we get $\hat x_\mu = x_\mu - ia^\alpha(L_{\mu\alpha}-L_{\alpha\mu})$ and we have 
\begin{equation}
[L_{\rho\sigma}, \hat{x}_{\nu}]=\eta_{\sigma\nu}\hat{x}_{\rho}-i\eta_{\sigma\nu}a^{\alpha}L_{\alpha\rho}+ia_{\sigma}L_{\rho\nu}-i\eta_{\rho\nu}a^{\alpha}L_{\alpha\sigma}+ia_{\rho}L_{\nu\sigma}.
\end{equation}
 If we define the Lorentz($SO(1,n-1)$) generators by $M_{\mu\nu}=L_{\mu\nu}-L_{\nu\mu}$, we get $\hat x_\mu = x_\mu - ia^\alpha M_{\mu\alpha}$ (where $[P_\mu, X_\nu]=-i\eta_{\mu\nu}$, $[X_\mu,X_\nu]=[P_\mu,P_\nu]=0$ and $X_\mu$ and $P_\mu$ transform vectorlike under $M_{\mu\nu}$) and we have 
\begin{equation}\label{kpoincare}
[M_{\rho\sigma}, \hat{x}_{\nu}]=\eta_{\sigma\nu}\hat{x}_{\rho}-\eta_{\rho\nu}\hat{x}_{\sigma}+ia_{\sigma}M_{\rho\nu}-ia_{\rho}M_{\nu\sigma}.
\end{equation}
Note that eq. \eqref{kpoincare} is expressed in terms of Lorentz generators $M_{\mu\nu}$. This implies the $\kappa$- Poincar\'e  algebra, which is the underlying symmetry of the differential algebra $\mathcal{C}_{4}$. It is important to emphasize that the differential algebra $\mathcal{C}_{4}$ is the only algebra of classical dimension compatible with the $\kappa$-Poincar\'e -Hopf algebra \eqref{kphalgebra} (see also subsection V.A.).

We note that this section can be generalized to the super-Hopf algebroid structure.


\section{Field theory}

The study of field theory over $\kappa$-Minkowski space is relevant for physics, since it may provide an interface between quantum gravity, NC geometry and their physical manifestations. Until today there is fairly large literature on $\kappa$-deformed field theory \cite{44c, ms11, klm00, klm00-, klry09, Govindarajan-2}, but all of these theories are very special, since they can be related to a specific realization or they are using the differential  calculus with one extra form.
Our goal is to give a framework for constructing field theory with differential calculus of classical dimension. In order to do this we need to introduce higher-degree forms, the Hodge-$*$ operation and an integral to define an action for the fields. 

In the usual undeformed differential geometry the higher-degree forms are defined via wedge product $\wedge$, but since one-forms in our approach generate a Grassmann algebra we have
$\boldsymbol\omega=\omega_{\alpha_1 ... \alpha_p}\d x^{\alpha_1}\wedge...\wedge\d x^{\alpha_p}\equiv \omega_{\alpha_1 ... \alpha_p}\xi^{\alpha_1}...\xi^{\alpha_p}\in\Omega^p.$ The Hodge-$*$ operation is defined as a mapping $*: \Omega^{p}\rightarrow\Omega^{n-p}$ via
\begin{equation}
\boldsymbol\alpha\wedge\boldsymbol\beta^{*}=\boldsymbol\alpha^{*}\wedge\boldsymbol\beta\equiv\boldsymbol\alpha\ (\boldsymbol\beta)^* = (\boldsymbol\alpha)^*\ \boldsymbol\beta=\alpha_{\mu_1 ...\mu_k}\beta^{\mu_1 ... \mu_k}\ \textbf{vol},
\end{equation}
where $\boldsymbol\alpha, \boldsymbol\beta\in\Omega^k$ and $\textbf{vol}=\d x^0 \wedge ...\wedge\d x^{n-1}\equiv\xi^0 ...\xi^{n-1}$ is the volume form. The integral is defined as a linear map 
\begin{equation}
\int: \Omega^{n}\rightarrow\mathbb{C}
\end{equation}
and it is closed in the sense that
$\int \d\boldsymbol\omega=0, \quad \forall \omega\in\Omega^n.$

\subsection{NC field theory}

So far we have established the connection between the usual methods of differential geometry and our algebraic approach. Now, we want to investigate the NC generalization and apply it to construction of NC field theories.

The higher degree forms in NC case are defined via
\begin{equation}
\hat{\boldsymbol\omega}=\hat{\omega}_{\alpha_1 ... \alpha_p}\hat{\xi}^{\alpha_1}...\hat{\xi}^{\alpha_p}\in\hat{\Omega}^p.
\end{equation}

The Hodge-$\hat{*}$ operation is defined as a mapping $\hat{*}: \hat{\Omega}^{p}\rightarrow\hat{\Omega}^{n-p}$ by
\begin{equation}\label{nchodge}
\hat{\boldsymbol\alpha}\ \ (\hat{\boldsymbol\beta})^{\hat{*}}=(\hat{\boldsymbol\alpha})^{\hat{*}}\  \hat{\boldsymbol\beta}\equiv \hat{\alpha}_{\mu_1 ... \mu_k}\hat{\beta}^{\mu_1 ... \mu_k}\ \hat{\text{vol}},
\end{equation}
where $\hat{\boldsymbol\alpha},\hat{\boldsymbol\beta}\in \hat{\Omega}^k $ and $\hat{\textbf{vol}}=\xi^0 ... \xi^{n-1}$ is the volume form. For $n=4$ we have 
\begin{equation}\begin{split}
&(1)^{\hat{*}}=\hat{\textbf{vol}}=\hat{\xi}^0 \hat{\xi}^1 \hat{\xi}^2 \hat{\xi}^3=\frac{1}{4! }\epsilon_{\mu\nu\rho\sigma}\hat{\xi}^{\mu}\hat{\xi}^{\nu}\hat{\xi}^{\rho}\hat{\xi}^{\sigma},\\
&(\hat{\xi}^{\mu})^{\hat{*}}=\frac{1}{3!}\epsilon^{\mu}_{\ \ \alpha_1 \alpha_{2}\alpha_{3}}\hat{\xi}^{\alpha_1}\hat{\xi}^{\alpha_2}\hat{\xi}^{\alpha_3},\\
&(\hat{\xi}^{\mu}\hat{\xi}^{\nu})^{\hat{*}}=\frac{1}{2!}\epsilon^{\mu\nu}_{\ \ \ \alpha_1 \alpha_2}\hat{\xi}^{\alpha_1}\hat{\xi}^{\alpha_2},\\
&(\hat{\xi}^{\mu}\hat{\xi}^{\nu}\hat{\xi}^{\rho})^{\hat{*}}=\epsilon^{\mu\nu\rho}_{\ \ \ \ \alpha}\hat{\xi}^{\alpha},\\
&(\hat{\xi}^{\mu}\hat{\xi}^{\nu}\hat{\xi}^{\rho}\hat{\xi}^{\sigma})^{\hat{*}}=\epsilon^{\mu\nu\rho\sigma}.
\end{split}\end{equation}

The integral is defined as a linear map
\begin{equation}\label{int}
\int: \hat{\Omega}^{n}\rightarrow\mathbb{C}.
\end{equation}
where the integral is closed in the sense that 
\begin{equation}\label{int1}
\int \hat{\d}\hat{\boldsymbol\omega}=0, \quad \forall \hat{\boldsymbol\omega}\in\hat{\Omega}^{n}.
\end{equation}
It is easy to see that for $n=4$ we have
\begin{equation}\label{int2}
\int \hat{\d}\hat{\boldsymbol\omega}=\int \hat{\xi}^{\mu}(\partial_{\mu}\blacktriangleright\hat{\omega}_{\alpha_1 \alpha_2 \alpha_3 \alpha_4})\hat{\xi}^{\alpha_1}\hat{\xi}^{\alpha_2}\hat{\xi}^{\alpha_3}\hat{\xi}^{\alpha_4}=\int (\partial_{\sigma}\Lambda^{\sigma}_{\ \mu}\blacktriangleright\hat{\omega}_{\alpha_1 \alpha_2 \alpha_3 \alpha_4})\hat{\xi}^{\mu}\hat{\xi}^{\alpha_1}\hat{\xi}^{\alpha_2}\hat{\xi}^{\alpha_3}\hat{\xi}^{\alpha_4}=0.
\end{equation}
At this level, the integral symbol defined by \eqref{int} and \eqref{int1} is just a formal notation. However, in the $\mathcal{C}_4$-case the integral is invariant under the action of $\kappa$-Poincar\'e algebra, so that the integral introduced here is the standard Lebesque integral applied to the functions which give a realization of the $\kappa$-Poincar\'e algebra through the $\star$-product \cite{44c, Durhuus, mercati}.

Now, we are ready to write an action $\hat{\textbf{S}}$ for a real NC scalar field $\hat{\phi}\in\hat{\mathcal{A}}$. We have
\begin{equation}
\hat{\textbf{S}}=\int \hat{\d}\hat{\phi}\ \ (\hat{\d}\hat{\phi})^{\hat{*}}+m^2\hat{\phi}\ \ (\hat{\phi})^{\hat{*}}.
\end{equation}
Since $\hat{\d}\hat{\phi}=(\partial_{\beta}\Lambda^{\beta}_{\ \alpha}\blacktriangleright\hat{\phi})\hat{\xi}^{\alpha}$ and using \eqref{nchodge} we have 
\begin{equation}\begin{split}
&\hat{\d}\hat{\phi}\ \ (\hat{\d}\hat{\phi})^{\hat{*}}=(\partial_{\beta}\Lambda^{\beta}_{\ \alpha}\blacktriangleright\hat{\phi})(\partial_{\rho}\Lambda^{\rho\alpha}\blacktriangleright\hat{\phi})\hat{\textbf{vol}}\\
&\hat{\phi}\ \ (\hat{\phi})^{\hat{*}}=\hat{\phi}\hat{\phi}\hat{\textbf{vol}},
\end{split}\end{equation}
so
\begin{equation}
\hat{\textbf{S}}=\int\left((\partial_{\beta}\Lambda^{\beta}_{\ \alpha}\blacktriangleright\hat{\phi})(\partial_{\rho}\Lambda^{\rho\alpha}\blacktriangleright\hat{\phi})+m^2\hat{\phi}\hat{\phi}\right)\hat{\textbf{vol}}.
\end{equation}
To find the equation of motion we impose $\delta\hat{\textbf{S}}=0$, that is
\begin{equation}\begin{split}
\delta\hat{\textbf{S}}&=\int \hat{\d}\delta\hat{\phi}\ \ (\hat{\d}\hat{\phi})^{\hat{*}}+ \hat{\d}\hat{\phi}\ \ (\hat{\d}\delta\hat{\phi})^{\hat{*}}+m^2\delta\hat{\phi}\ \ (\hat{\phi})^{\hat{*}}+m^2\hat{\phi}\ \ (\delta\hat{\phi})^{\hat{*}}   \\
&=\int -\delta\hat{\phi}\ \ [\hat{\d}(\hat{\d}\hat{\phi})^{\hat{*}}]  +(\hat{\d}\hat{\phi})^{\hat{*}}\ \ (\hat{\d}\delta\hat{\phi})+m^2\delta\hat{\phi}\ \ (\hat{\phi})^{\hat{*}}+m^2(\hat{\phi})^{\hat{*}}\ \ \delta\hat{\phi}   \\
&=\int \delta\hat{\phi}\ \ \left[-\hat{\d}(\hat{\d}\hat{\phi})^{\hat{*}}+m^2(\hat{\phi})^{\hat{*}}\right]+\left[-\hat{\d}(\hat{\d}\hat{\phi})^{\hat{*}}+m^2(\hat{\phi})^{\hat{*}}\right]\ \ \delta\hat{\phi}
\end{split}\end{equation}
which leads to
\begin{equation}\label{nceom}
\left[\hat{\d}(\hat{\d}\hat{\phi})^{\hat{*}}\right]^{\hat{*}}=m^2\hat{\phi}
\end{equation}
where we used 
\begin{equation}\begin{split}
&\int \hat{\d}[\delta\hat{\phi}\ \ (\hat{\d}\hat{\phi})^{*}]=0=\int \hat{\d}\delta\hat{\phi}\ \ (\hat{\d}\hat{\phi})^{\hat{*}} + \int \delta\hat{\phi}\ \ [\hat{\d}(\hat{\d}\hat{\phi})^{\hat{*}}],\\
&\int \hat{\d}[\hat{\d}\hat{\phi}\ \ (\delta\hat{\phi})^{\hat{*}}] =0=\int \hat{\d}(\hat{\d}\hat{\phi})^{\hat{*}}\ \ \delta\hat{\phi} +\int  \hat{\d}\hat{\phi}\ \ (\hat{\d}\delta\hat{\phi})^{\hat{*}}, \\
&((\hat{\phi})^{\hat{*}})^{\hat{*}}=\hat{\phi}.
\end{split}\end{equation}
Eq. \eqref{nceom} represents the NC generalization of the Klein-Gordon equation.  Let us investigate the l.h.s. of eq.\eqref{nceom}. We have
\begin{equation}\begin{split}
\left[\hat{\d}(\hat{\d}\hat{\phi})^{\hat{*}}\right]^{\hat{*}}&=\left[\hat{\d}\left((\partial_{\beta}\Lambda^{\beta}_{\ \alpha}\blacktriangleright\hat{\phi})\hat{\xi}^{\alpha}\right)^{\hat{*}}\right]^{\hat{*}}\\
&=\left[\hat{\d}\left((\partial_{\beta}\Lambda^{\beta}_{\ \alpha}\blacktriangleright\hat{\phi})\frac{1}{3!}\epsilon^{\alpha}_{\ \ \rho_1 \rho_2 \rho_3}\hat{\xi}^{\rho_1}\hat{\xi}^{\rho_2}\hat{\xi}^{\rho_3}\right)\right]^{\hat{*}}\\
&=\left[(\partial_{\gamma}\Lambda^{\gamma}_{\ \delta}\partial_{\beta}\Lambda^{\beta}_{\ \alpha}\blacktriangleright\hat{\phi})\frac{1}{3!}\epsilon^{\alpha}_{\ \ \rho_1 \rho_2 \rho_3}\hat{\xi}^{\delta}\hat{\xi}^{\rho_1}\hat{\xi}^{\rho_2}\hat{\xi}^{\rho_3}\right]^{\hat{*}}\\
&=\left[(\partial_{\gamma}\Lambda^{\gamma}_{\ \delta}\partial_{\beta}\Lambda^{\beta\delta}\blacktriangleright\hat{\phi})\hat{\text{vol}}\right]^{\hat{*}}\\
&=\partial_{\gamma}\Lambda^{\gamma}_{\ \delta}\partial_{\beta}\Lambda^{\beta\delta}\blacktriangleright\hat{\phi}.
\end{split}\end{equation}
So, for the equation of motion we have
\begin{equation}
\partial_{\alpha}\partial_{\beta}\Lambda^{\alpha}_{\ \sigma}\Lambda^{\beta\sigma}\blacktriangleright\hat{\phi}-m^2\hat{\phi}=0.
\end{equation}

\subsection{Dispersion relations}

We can look at some specific examples for $\Lambda^{\alpha}_{\ \sigma}$ and derive the NC dispersion relations. For example for $\mathcal{S}_1$ we have $\Lambda_{\mu\alpha}=\eta_{\mu\alpha}Z^{-1}$ so for the NC Klein-Gordon equation we get
\begin{equation}
\mathcal{S}_1: \quad (\partial^R)^2 Z^{-2}\blacktriangleright\hat{\phi}-m^2\hat{\phi}=0,
\end{equation}
and for $\mathcal{S}_2$, where $\Lambda_{\mu\alpha}=\eta_{\mu\alpha}+ia_{\mu}\partial^{L}_{\alpha}Z$ we get
\begin{equation}
\mathcal{S}_2: \quad (\partial^L)^2 Z^{2}\blacktriangleright\hat{\phi}-m^2\hat{\phi}=0
\end{equation}
where $Z$ is given in \eqref{Za}.
We see that the main new NC feature is the modification of dispersion relations
\begin{equation}\begin{split}\label{disp}
&\mathcal{S}_1: \quad E^2-\vec{p}^2=(mZ)^2, \quad Z=1-ap\\
&\mathcal{S}_2: \quad E^2-\vec{p}^2=\left(\frac{m}{Z}\right)^2, \quad  Z=\frac{1}{1+ap}
\end{split}\end{equation}
 It is interesting to examine $\mathcal{C}_4$, where $\Lambda_{\mu\nu}=\eta_{\mu\nu}+ia_{\mu}D_{\nu}-ia_{\nu}D_{\mu}+\frac{1}{2}a_{\mu}a_{\nu}D^2 Z$ leads to
\begin{equation}
D^2\blacktriangleright\hat{\phi}-m^2\hat{\phi}=0.
\end{equation}
We see that in the case of $\mathcal{C}_4$ we get $D^2=\Box$ ($a^2=0$), that is the Casimir operator of the Poincar\'e algebra. This was expected, since $\mathcal{C}_4$ is compatible with the Poincar\'e algebra, and only the coalgebraic sector is deformed.

The dispersion relations we obtain here are in general different from the ones usually found in the literature.  It is usual to investigate the operator $\Box$ in different realizations. Operator $\Box$   is the generalized d'Alembert operator and it commutes with the Lorentz generators $M_{\mu\nu}$ and translation generators $p_{\mu}\equiv-i\partial_{\mu}$ for any realization of $M_{\mu\nu}$ and $p_{\mu}$, which means that it is a Casimir operator of the deformed Poincar\'e algebra and we can assign to it an invariant mass $m$ in the following way
\begin{equation}\label{laplace}
\Box=m^2 .
\end{equation}
The generalized d'Alembert operator is given by:
\begin{equation}
\Box=\frac2{a^2}\left(
1-\sqrt{1-a^2D^2}
\right).
\end{equation}
We will use the following notation for the partial derivatives in each realization:
\begin{equation}\begin{split}
&\text{right covariant}: \quad \partial_{\mu}\equiv\partial^{R}_{\mu}\\
&\text{left covariant}: \quad \partial_{\mu}\equiv\partial^{L}_{\mu}\\
&\text{Magueijo-Smolin}: \quad \partial_{\mu}\equiv\partial^{MS}_{\mu}\\
\end{split}\end{equation}
where the superscripts $R, L, MS$ denote the \textsl{right covariant}, \textsl{left covariant} and \textsl{Magueijo-Smolin} realizations respectively, and $D_{\mu}$ stands for the natural realization or classical basis (for more details on realizations see \cite{Meljanac-3, ms06, ms11, 90} ). All four realizations are connected via similarity transformations (see Section III and \cite{algebroid, Meljanac-3, ms06, ms11, 90}) and they are explicitly connected in the following way
\begin{equation}\begin{split}\label{veze}
&\partial^{R}_{\mu}=\partial^{L}_{\mu}Z,\\
&D_{\mu}=\partial^{L}_{\mu}+\frac{ia_{\mu}}{2}\Box,\\
&D_{\mu}=\partial^{MS}_{\mu}Z^{-1},
\end{split}\end{equation}
where $D^2=\Box\left(1-\frac{a^2}{4}\Box\right)$ and Z is the shift operator given by
\begin{equation}\label{Z}
Z=1+ia\partial^R=\frac{1}{1-ia\partial^L}=\sqrt{(1+ia\partial^{MS})^2+a^2(\partial^{MS})^2}=\frac{1}{-iaD+\sqrt{1-a^2D^2}}.
\end{equation}
Using \eqref{veze}, \eqref{Z}, \eqref{laplace} and $p_{\mu}=-i\partial_{\mu}$ we get the following dispersion relations
\begin{equation}\begin{split}
&\text{right covariant}: \quad (p^R)^2=-m^2Z=-m^2(1-ap^R)\\
&\text{left covariant}: \quad (p^L)^2=-m^2Z^{-1}=-m^2(1+ap^L)\\
&\text{Magueijo-Smolin}: \quad (p^{MS})^2=-m_{eff}^2(1-ap^{MS})^2\\
\end{split}\end{equation}
where $m^2_{eff}=\frac{m^2\left(1+\frac{a^2}{4}m^2\right)}{1+a^2m^2\left(1+\frac{a^2}{4}m^2\right)}$. For the time-like deformations of Minkowski space \eqref{kappa} we get
\begin{equation}\begin{split}
&\text{right covariant}: \quad E^2-\vec{p}^2=m^2(1-a_{0}E)\\
&\text{left covariant}: \quad E^2-\vec{p}^2=m^2(1+a_{0}E)\\
&\text{Magueijo-Smolin}: \quad E^2-\vec{p}^2=m_{eff}^2(1-a_{0}E)^2\\
\end{split}\end{equation}
where we used $a_{\mu}=(a_0,\vec{0})$ and $p^{\mu}=(E,\vec{p})$, where $a_0=\frac1\kappa$. The difference between these two approaches is mainly in the fact that the operator in the NC Klein-Gordon equation \eqref{nceom} is not related to the d'Alambertian invariant under the $\kappa$-Poincar\'e algebra, but rather an operator invariant under certain $\kappa$-deformation of the $\mathfrak{igl}(n)$ algebra.

\section{Final remarks}

The framework developed so far will enable us to describe and further investigate the physics of other field theories.
For example we can also examine $\kappa$-deformed electrodynamics via 
\begin{equation}
\hat{\textbf{S}}=-\frac{1}{4}\int \hat{\textbf{F}}\ \ (\hat{\textbf{F}})^{\hat{*}},
\end{equation}
where $\hat{\textbf{F}}=\hat{\d}\hat{\textbf{A}}$. The equations of motion are given by $\delta\hat{\textbf{S}}=0$, that is
\begin{equation}
\hat{\d}(\hat{\d}\hat{\textbf{A}})^{\hat{*}}=0, \quad \Leftrightarrow \quad \hat{\d}(\hat{\textbf{F}})^{\hat{*}}=0.
\end{equation}
The NC version of Bianchi identity also holds $\hat{\d}\hat{\textbf{F}}=\hat{\d}(\hat{\d}\hat{\textbf{A}})=\hat{\eta}^2\blacktriangleright\hat{\textbf{A}}=0$.

Similarly, for the NC generalization of the Abelian Chern-Simons theory in $n=3$ the action is given by
\begin{equation}
\hat{\textbf{S}}=k\int \hat{\textbf{A}}\ \ \hat{\d}\hat{\textbf{A}}
\end{equation}
and the corresponding equation of motion is
\begin{equation}
\hat{\d}\hat{\textbf{A}}=0 \quad \Leftrightarrow \quad \hat{\textbf{F}}=0.
\end{equation}
Further investigations of field theories in this NC setting will be reported elsewhere.

So far we have analyzed the NC version of the free classical field theory. The new physical contribution is the modification of dispersion relations \eqref{disp}. The next step would be to investigate  the interacting classical field theory, by adding additional terms in the action, and then analyze the case of quantum field theory. In order to do this , we have to investigate the $R$-matrix which would modify the quantization procedure, that is we have to modify the algebra of creation and annihilation operators via
\begin{equation}
\phi(x)\otimes\phi(y)-R\phi(y)\otimes\phi(x)=0
\end{equation}
which will modify the usual spin-statistics relations of free bosons at Planck scale. The $R$-matrix is defined by the twist operator $R=\tilde{\mathcal{F}}\mathcal{F}^{-1}$. The issue of star-product quantization is still under investigation. The idea is that the $R$-matrix will enable us to define particle statistics and to properly quantize fields, while the twist operator will provide star-product, which is crucial for writing the action in terms of commutative variables, which will in the end enable us to derive the Feynman rules and see the NC correction to the propagator and vertex. 

The general plan for future work is to study differential calculus on $\kappa$-Minkowski space and related concepts (e.g. vector fields, Lie derivative, exterior derivative, Hodge star-operator and integration) using two approaches: the twist deformation method \cite{49} and the method of realizations \cite{23}. Using a Drinfeld twist operator (or a twist in the Hopf algebroid setting \cite{algebroid}) one can introduce noncommutativity via star-product on the algebra $\mathcal{A}$ of complex-valued functions on the classical Minkowski space. We want to study the induced deformations of $\mathcal{A}$-modules of vector fields and one-forms on $\kappa$-Minkowski space as well as deformations of the related morphisms. An alternative approach to deformation of the differential algebra is based on realizations of coordinates and one-forms on $\kappa$-Minkowski space as power series in a formal parameter with coefficients in the super-Heisenberg algebra.

In this paper we have not discussed the issue of NC vector fields. Usually, in NC geometry \cite{47, DV, connes} vector fields are replaced with derivations of the NC algebra. We prefer to define NC vector fields  via $\hat{V}=\hat{V}^{\mu}\partial_{\mu}$, where 
$\hat{V}^{\mu}\in\hat{\mathcal{A}}$ and $\hat{V}$ satisfies a deformed Leibniz rule because of \eqref{unicop}.   This choice would be the natural generalization of the usual vector fields in the commutative case and it is easy to see that this definition reduces to the usual definition of vector field on commutative manifold after $a_{\mu}\rightarrow 0$. The duality between NC differential forms and NC vector fields is of immense importance and is a work in progress. 

The existence  of Cartan operations, i.e. exterior derivative, Lie derivative and inner derivative and their mutual commutation relations is under our current investigation. Namely, the exterior derivative is defined by \eqref{d}, while the inner derivative is related to the action of the generator $q_{\mu}$. The Lie derivative in our approach was first discuss in \cite{EPJC} (see the text after eq. (11) in \cite{EPJC}). 

\bigskip

\noindent{\bf Acknowledgment}\\
Authors would like to thank F. Toppan, J.C. Wallet, J. Lukierski, A. Borowiec  and A. Pachol for valuable discussions and  comments.  R. \v S. thanks the Rudjer Bo\v skovi\'c Institute for hospitality during her visit. T. J. thanks LPT-Orsay for the hospitality. This work has been fully supported by Croatian Science Foundation under the project (IP-2014-09-9582).

\appendix
\section{Universal formula for Lie algebras}
Let us define a Lie algebra generated by $\hat x_\mu$
\begin{equation}\label{1}
[\hat x_\mu, \hat x_\nu]=iC_{\mu\nu}{}^\lambda \hat x_\lambda.
\end{equation}
The Heisenberg algebra is defined by
\begin{equation}\label{2}
[x_\mu,x_\nu]=[\partial_\mu,\partial_\nu]=0,\quad [\partial_\mu,x_\nu]=\eta_{\mu\nu}.
\end{equation}
One can find the realization of the Lie algebra \eqref{1} in terms the of the completion of the Heisenberg algebra \eqref{2} in first order in $C_{\mu\nu\lambda}$, which is related to the Weyl ordering \cite{Meljanac-3}:
\begin{equation}
\hat x_\mu=x_\mu+\frac{i}2x^\alpha C_{\mu\beta\alpha}\partial^\beta+O(C^2).
\end{equation}
In \cite{Durov} is the construction of the \textsl{universal formula} for $\hat x_\mu$ valid in all orders in $C_{\mu\nu\alpha}$
\begin{equation}
\hat x_\mu=x^\alpha\left(\frac{\EuScript C}{1-\text e^{-\EuScript C}}\right)_{\mu\alpha}
\end{equation}
where $\EuScript C_{\mu\nu}=iC_{\mu\alpha\nu}(\partial^W)^\alpha$. This realization corresponds to the Weyl ordering. 

We can introduce another Lie algebra generated by $\hat y_\mu$ and satisfying \cite{algebroid}
\begin{equation}\label{l2}
[\hat y_\mu, \hat y_\nu]=-iC_{\mu\nu}{}^\lambda\hat y_\lambda.
\end{equation}
Now, the realization of $\hat y_\mu$ is also given by the \textsl{universal formula}
\begin{equation}
\hat y_\mu=x^\alpha\left(\frac{\EuScript C}{\text e^{\EuScript C}-1}\right)_{\mu\alpha}.
\end{equation}

There is an antisomorphism between Lie algebras \eqref{1} and \eqref{l2} defined by $\hat{x}\mapsto\hat{y}$. In the Heisenberg algebra there exist an identity among the realization of $\hat{x}$ and $\hat{y}$ given by
\begin{equation}
\hat x_\mu=\hat y^\alpha O_{\mu\alpha}
\end{equation}
where $O_{\mu\nu}$ is given by
\begin{equation}\label{O}
O_{\mu\nu}=\left[\text e^{\EuScript C}\right]_{\mu\nu},
\end{equation}
and it can be shown that $\hat x_\mu$ and $\hat y^\alpha$ commute
\begin{equation}
[\hat x_\mu, \hat y^\alpha]=0
\end{equation}

We restrict ourselves to the special case when
\begin{equation}\label{c}
C_{\mu\nu\lambda}=a_{\mu}\eta_{\nu\lambda}-a_{\nu}\eta_{\mu\lambda}.
\end{equation}
It follows that combining \eqref{O} and \eqref{c} we have (we shall omit the label $W$ for the sake of simplicity)
\begin{equation}\begin{split}
O_{\mu\nu}&=\left[\text{e}^{\EuScript{C}}\right]_{\mu\nu}=\text{e}^{-i\eta_{\mu\nu}a\partial+ia_{\mu}\partial_{\nu}}=\text{e}^{-i\eta_{\mu\nu}a\partial}\text{e}^{ia_{\mu}\partial_{\nu}}=\eta_{\mu\alpha}\text{e}^{-ia\partial}\left(\eta^{\alpha}_{\ \nu}+ia^{\alpha}\partial_{\nu}\frac{1-\text{e}^{-ia\partial}}{ia\partial}\right)\\
&=\eta_{\mu\nu}Z^{-1}+ia_{\mu}(\partial^{L})_{\nu},
\end{split}\end{equation}
where $a\partial \equiv a_\alpha\partial^\alpha$ and we used $Z=\text{e}^{ia\partial}$ and $(\partial^{L})_{\nu}=\partial_{\nu}\frac{1-\text{e}^{-ia\partial}}{ia\partial}$.

Note that $\hat{y}_{\mu}$ can be interpreted as the right multiplication
\begin{equation}
\hat y_\mu \blacktriangleright \hat f=\hat f\hat x_\mu, \quad \forall \hat f \in\hat{\mathcal A}
\end{equation}
We also have \eqref{xffx}, \eqref{Oxcomm} and
\begin{equation}
[O_{\mu\nu}, \hat y_\lambda]=i(a_\mu\eta_{\lambda\nu}- a_\lambda O_{\mu\nu}),
\end{equation}
and similarly for $[O^{-1}]_{\mu\nu}$. 

\section{Explicit calculation of $\Lambda_{\mu\nu}$}

Here we will explicitly illustrate the above construction for the special cases of covariant differential algebras introduced in Section II.

\bigskip

\textsl{Differential algebra $\mathcal S_1$}:

\bigskip

$\mathcal S_1$ is defined by 
\begin{equation}
K_{\mu\nu\alpha}=-\eta_{\mu\alpha}a_\nu
\end{equation}
and corresponds to the \textsl{right covariant ordering}. Using \eqref{lambdaexp} we have
\begin{equation}
\Lambda_{\mu\alpha}=[\text{e}^{\rho}]_{\mu\alpha}=\eta_{\mu\alpha}Z^{-1},
\end{equation}
where $\rho_{\mu\alpha}=-i\eta_{\mu\alpha}a\partial^W$.
Obviously $[\Lambda^{-1}]_{\mu\alpha}=\eta_{\mu\alpha}Z$, so using \eqref{unicop} and \eqref{deltaq} we get
\begin{equation}
\Delta \partial^R_\mu=\partial^R_\mu\otimes 1+Z\otimes\partial^R_\mu, \quad \Delta q_\mu=q_\mu\otimes 1+ (-)^{N_1}Z\otimes q_\mu,
\end{equation}
where we used $Z=\text e^{ia\partial^W}=1+ia\partial^R$.

\bigskip

\textsl{Differential algebra $\mathcal S_2$}:

\bigskip

$\mathcal S_2$ is defined by 
\begin{equation}
K_{\mu\nu\alpha}=\eta_{\nu\alpha}a_\mu
\end{equation}
and corresponds to the \textsl{left covariant ordering}. Using \eqref{lambdaexp} we have
\begin{equation}
\Lambda_{\mu\alpha}=[\text e^\rho]_{\mu\alpha}
\end{equation}
where $\rho_{\mu\alpha}=ia_\mu\partial^W_\alpha$. Note that $\rho^n=\rho A_W^{n-1}$, where $A_W=ia\partial^W$, so $\text e^\rho=1+\rho\frac{\text e^{A_W}-1}{A_W}$, that is
\begin{equation}
\Lambda_{\mu\alpha}=\eta_{\mu\alpha}+ia_\mu\partial^W_\alpha\frac{\text e^{A_W}-1}{A_W}=\eta_{\mu\alpha}+ia_\mu\partial^L_\alpha Z.
\end{equation}
Since $[\Lambda^{-1}]_{\mu\alpha}=\eta_{\mu\alpha}-ia_\mu\partial^L_\alpha$, using \eqref{unicop} and \eqref{deltaq} we get
\begin{equation}
\Delta \partial^L_\mu=\partial^L_\mu\otimes Z^{-1}+1\otimes\partial^L_\mu, \quad \Delta q_\mu=\Delta_0q_\mu+(-)^{N_1}\partial^L_\mu\otimes (-iaq),
\end{equation}
where we used $Z=\text e^{ia\partial^W}=\frac1{1-ia\partial^L}$. 

\bigskip

\textsl{Differential algebra $\mathcal S_3$}:

\bigskip

$\mathcal S_3$ is defined by 
\begin{equation}
K_{\mu\nu\alpha}=-\eta_{\mu\nu}a_\alpha-\eta_{\mu\alpha}a_\nu
\end{equation}
and corresponds to the \textsl{Magueijo-Smolin realization}. Using \eqref{lambdaexp} we have
\begin{equation}
\Lambda_{\mu\alpha}=\text{exp}(-i\eta_{\mu\alpha}a\partial^W)=\text e^{-i\eta_{\mu\alpha}a\partial^W}\text e^{-ia_\alpha\partial^W_\mu}=\eta_{\mu\alpha}Z^{-1}-ia_\alpha\partial^L_\mu Z^{-1},
\end{equation}
where we used $Z=\text e^{ia\partial^W}=\sqrt{(1+ia\partial^{MS})^2+a^2(\partial^{MS})^2}$ and $\partial^{MS}_\mu=\partial^L_\mu Z+\frac{ia_\mu}2(\partial^L)^2 Z^2$.
Since $[\Lambda^{-1}]_{\mu\alpha}=\eta_{\mu\alpha}Z+ia_\alpha\partial^L_\mu Z^2$,  using \eqref{unicop} and \eqref{deltaq} we get
\begin{equation}
\Delta \partial^{MS}_\mu=\partial^{MS}_\mu\otimes 1+Z\otimes\partial^{MS}_\mu+ia_\mu Z^2(\partial^L)^\alpha\otimes\partial^{MS}_\alpha, \quad \Delta q_\mu=q_\mu\otimes 1+ (-)^{N_1}(Z\otimes q_\mu+ia_\mu Z^2(\partial^L)^\alpha\otimes q_\alpha).
\end{equation}

\bigskip

\textsl{Differential algebra $\mathcal{C}_{4}$}:

\bigskip

$\mathcal{C}_{4}$ is defined by 
\begin{equation}
K_{\mu\nu\alpha}=-\eta_{\mu\nu}a_\alpha+\eta_{\nu\alpha}a_\mu
\end{equation}
and corresponds to the \textsl{natural realization} (or classical basis) for $a^2=0$. Using \eqref{lambdaexp} we have
\begin{equation}
\Lambda_{\mu\nu}=\eta_{\mu\nu}+ia_\mu\partial^R_\nu-ia_\nu D_\mu
\end{equation}
where we used $D_\mu=\partial^L_\mu+ia_\mu(\partial^W)^2\frac{\cosh A_D -1}{A_D^2}$ and $\partial^L_\mu=\partial^R_\mu Z^{-1}=\partial^W_\mu\frac{1-\text e^{-A_D}}{A_D}$, where $A_D=iaD$. Since $[\Lambda^{-1}]_{\mu\nu}=\eta_{\mu\nu}-ia_{\mu}D_{\nu}+ia_{\nu}(D_{\mu}-\frac{i}{2}a_{\mu}D^2)Z$,   for the coproducts we have
\begin{equation}
\Delta D_{\mu}=\left[-ia^\alpha D_\mu + ia_\mu\left(D^\alpha - \frac12 ia^\alpha D^2 \right) Z \right] \otimes D_\alpha + \Delta_0 D_\mu.
\end{equation}

\end{document}